\documentclass[graybox]{svmult}

\usepackage{mathptmx}       
\usepackage{helvet}         
\usepackage{courier}        
\usepackage{type1cm}        
\usepackage{makeidx}         
\usepackage{graphicx}        
\usepackage{multicol}        
\usepackage[bottom]{footmisc}

\usepackage{amsfonts}
\usepackage{amsmath}
\usepackage{amssymb}
\usepackage{bm}
\usepackage{hyperref}
\usepackage{color}

\allowdisplaybreaks

\newcommand{\R}{\mathbb{R}}
\newcommand{\Z}{\mathbb{Z}}
\newcommand{\N}{\mathbb{N}}
\newcommand{\C}{\mathbb{C}}
\newcommand{\ui}{\mathrm{i}}
\newcommand{\ud}{\mathrm{d}}
\newcommand{\uD}{\mathrm{D}}
\newcommand{\FP}{\mathop{\mathrm{FP}}_{B=0}}
\newcommand{\ab}{^{\alpha\beta}}
\newcommand{\lab}{_{\alpha\beta}}

\makeindex             

\begin{document}

\title*{Post-Newtonian theory and the two-body problem}
\titlerunning{Post-Newtonian theory and the two-body problem} 
\author{Luc Blanchet}
\institute{
Luc Blanchet \at
${\mathcal{G}}{\R}\varepsilon{\C}{\mathcal{O}}$, Institut d'Astrophysique de Paris --- C.N.R.S. \& Universit\'e Pierre et Marie Curie, 98$^{\mathrm{bis}}$ boulevard Arago, 75014 Paris, France. \email{blanchet@iap.fr}
}
%
%
\maketitle

\abstract{Reliable predictions of general relativity theory are extracted using approximation methods. Among these, the powerful post-Newtonian approximation provides us with our best insights into the problems of motion and gravitational radiation of systems of compact objects. This approximation has reached an impressive mature status, because of important progress regarding its theoretical foundations, and the successful construction of templates of gravitational waves emitted by inspiralling compact binaries. The post-Newtonian predictions are routinely used for searching and analyzing the very weak signals of gravitational waves in current generations of detectors. High-accuracy comparisons with the results of numerical simulations for the merger and ring-down of binary black holes are going on. In this article we give an overview on the general formulation of the post-Newtonian approximation and present up-to-date results for the templates of compact binary inspiral.} 

\section{Introduction}\label{sec1}

Although relativists admire the mathematical coherence --- and therefore beauty --- of Einstein's general relativity, this theory is not easy to manage when drawing firm predictions for the outcome of laboratory experiments and astronomical observations. Indeed only few exact solutions of the Einstein field equations are known, and one is obliged in most cases to resort to approximation methods. It is beyond question that approximation methods in general relativity do work, and in some cases with some incredible precision. Many of the great successes of general relativity were in fact obtained using approximation methods. However because of the complexity of the field equations, such methods become awfully intricate at high approximation orders. On the other hand it is difficult to set up a formalism in which the approximation method would be perfectly well-defined and based on clear premises. Sometimes it is impossible to relate the approximation method to the exact framework of the theory. In this case the only thing one can do is to rely on the approximation method as the only representation of real phenomena, and to discover ``empirically'' that the approximation works well.

The most important approximation scheme in general relativity is the post-Newtonian expansion, which can be viewed as an expansion when the speed of light $c$ tends to infinity, and is physically valid under the assumptions of weak gravitational field inside the source and of slow internal motion. The post-Newtonian approximation makes sense only in the \textit{near-zone} of an isolated matter source, defined as $r\ll \lambda$, where $\lambda\equiv c\,T$ is the wavelength of the emitted gravitational radiation, with $T$ being a characteristic time scale of variation of the source. The approximation has been formalized in the early days of general relativity by Einstein \cite{E16}, De Sitter \cite{dS16a, dS16b}, Lorentz \& Droste \cite{LD17}. It was subsequently developed notably by Einstein, Infeld \& Hoffmann \cite{EIH}, Fock \cite{Fock39, Fock}, Plebanski \& Bazanski \cite{PB59}, Chandrasekhar and collaborators \cite{C65, CN69, CE70}, Ehlers and his school \cite{Ehl80, K80a, K80b}, Papapetrou and coworkers \cite{Papa51, PapaL81}.

Several long-standing problems with the post-Newtonian approximation for general isolated slowly-moving systems have hindered progress untill recently. At high post-Newtonian ordners some \textit{divergent} Poisson-type integrals appear, casting a doubt on the physical soundness of the approximation. Linked to that, the domain of validity of the post-Newtonian approximation is limited to the near-zone of the source, making it \textit{a priori} difficult to incorporate into the scheme the condition of no-incoming radiation, to be imposed at past null infinity from an isolated source. In addition, from a mathematical point of view, we do not know what the ``reliability'' of the post-Newtonian series is, \textit{i.e.} if it comes from the Taylor expansion of a family of exact solutions.

The post-Newtonian approximation gives wonderful answers to the problems of motion and gravitational radiation, two of general relativity's corner stones. Three crucial applications are:
\begin{enumerate}
\item The motion of $N$ point-like objects at the first post-Newtonian level \cite{EIH} \textit{i.e.} 1PN,\footnote{As usual, we refer to $n$PN as the order equivalent to terms $\sim (v/c)^{2n}$ in the equations of motion beyond the Newtonian acceleration, and in the asymptotic waveform beyond the Einstein quadrupole formula, where $v$ denotes the binary's orbital velocity and $c$ is the speed of light.} is taken into account to describe the Solar System dynamics (motion of the centers of mass of planets); 
\item The gravitational radiation-reaction force, which appears in the equations of motion at the 2.5PN order \cite{DD81a, DD81b, D83, D83houches}, has been experimentally verified by the observation of the secular acceleration of the orbital motion of the Hulse-Taylor binary pulsar PSR~1913+16 \cite{TFMc79, TW82, T93};
\item The analysis of gravitational waves emitted by inspiralling compact binaries --- two neutron stars or black holes driven into coalescence by emission of gravitational radiation --- necessitate the prior knowledge of the equations of motion and radiation field up to high post-Newtonian order.
\end{enumerate}
Strategies to detect and analyze the very weak signals from inspiralling compact binaries involve matched filtering of a set of accurate theoretical template waveforms against the output of the detectors. Measurement-accuracy analyses have shown that, in order to get sufficiently accurate theoretical templates, one must at least include conservative post-Newtonian effects up to the 3PN level, and radiation-reaction effects up to 5.5PN order, \textit{i.e.} 3PN beyond the leading order of radiation reaction which is 2.5PN \cite{3mn, CFPS93, FCh93, CF94, TNaka94, P95}.

The appropriate description of inspiralling compact binaries is by two structureless \textit{point-particles}, characterized solely by their masses $m_1$ or $m_2$ (and possibly their spins). Indeed, most of the non-gravitational effects usually plaguing the dynamics of binary star systems, such as the effects of a magnetic field, of an interstellar medium, the influence of the internal structure of the compact bodies, are dominated by purely gravitational effects. Inspiralling compact binaries are very clean systems which can essentially be described in pure general relativity. 

Although point-particles are ill-defined in the exact theory, they are admissible in post-Newtonian approximations. Furthermore the model of point particles can be pushed to high post-Newtonian order, where an \textit{a priori} more realistic model involving the internal structure of compact bodies would fail through becoming intractable. However there is an important worry: a process of regularization to deal with the infinite self-field of point particles is crucially needed. The regularization should be carefully defined to be implemented at high orders. It should hopefully be followed by a renormalization.

The orbit of inspiralling compact binaries can be considered to be circular, apart from the gradual inspiral, with an excellent approximation. Indeed, gravitational radiation reaction forces tend to circularize rapidly the orbital motion. At each instant during the gradual inspiral, the eccentricity $e$ of the orbit is related to the instantaneous frequency $\omega\equiv 2\pi/P$ by \cite{Pe64}
\begin{equation}\label{e2}
e^2 \simeq \mathrm{const} \,\omega^{-19/9}\qquad\text{(for $e\ll 1$)}\,.
\end{equation}
For instance one can check that the eccentricity of a system like the binary pulsar PSR~1913+16 will be $e\simeq 5\,10^{-6}$ when the gravitational waves become visible by the detectors, \textit{i.e.} when the signal frequency after its long chirp reaches $f\equiv\omega/\pi\simeq 10\,\mathrm{Hz}$. Only systems formed very late, near their final coalescence, could have a non-negligible residual eccentricity.

The intrinsic rotations or spins of the compact bodies could play an important role, yielding some relativistic spin-orbit and spin-spin couplings, both in the binary's equations of motion and radiation field. The spin of a rotating body is of the order of $S\sim m\,a\,v_\mathrm{spin}$, where $m$ and $a$ denote the mass and typical size of the body, and where $v_\mathrm{spin}$ represents the velocity of the body's surface. In the case of compact bodies we have $a\sim G\,m/c^2$, and for maximal rotation $v_\mathrm{spin}\sim c$; for such objects the magnitude of the spin is roughly $S\sim G\,m^2/c$. It is thus customary to introduce a dimensionless spin parameter, generally denoted by $\chi$, defined by
\begin{equation}\label{chi}
S = \frac{G\,m^2}{c}\,\chi\,.
\end{equation}
We have $\chi\leq 1$ for black-holes, and $\chi\lesssim 0.63-0.74$ for neutron stars (depending on the equation of state of nuclear matter inside the neutron star). For binary pulsars such as PSR~1913+16, we have $\chi\lesssim 5\,10^{-3}$. Considering models of evolution of observed binary pulsar systems when they become close to the coalescence we expect that the spins will make a negligible contribution to the accumulated phase in this case. However, astrophysical observations suggest that black holes can have non-negligible spins, due to spin up driven by accretion from a companion during some earlier phase of the binary evolution. For a few black holes surrounded by matter, observations indicate a significant intrinsic angular momentum and the spin may even be close to its maximal value. However, very little is known about the black-hole spin magnitudes in binary systems. For black holes rotating near the maximal value the templates of gravitational waves need to take into account the effects of spins, both for a successful detection and an accurate parameter estimation.

We devote Part A of this article, including Sections \ref{sec2} to \ref{sec7}, to a general overview of the formulation of the post-Newtonian approximation for isolated sources. We emphasize that the approximation can be carried out up to any post-Newtonian order, without the aforesaid problem of divergences. The main technique used is the matching of asymptotic expansions which permits to obtain a complete post-Newtonian solution incorporing the correct boundary conditions at infinity. Then Part B, \textit{i.e.} Sections \ref{sec8} to \ref{sec13}, will deal with the application to systems composed of compact objects. The subtle issues linked with the self-field regularization of point-particles are discussed in some details. The results for the two-body equations of motion and radiation field at the state-of-the-art 3PN level are presented in the case of circular orbits appropriate to inspiralling compact binaries. We put the accent on the description of spinning particles and particularly on the spin-orbit coupling effect on the binary's internal energy and gravitational-wave flux.

\bigskip\bigskip
\centerline{\large{\bf{A~ Post-Newtonian formalism}}}
\vspace{-0.4cm}
\section{Einstein field equations}
\label{sec2}

General relativity is based on two independent tenets, the first one concerned with the dynamics of the gravitational field, the second one dealing with the coupling of all the matter fields with the gravitational field. Accordingly the action of general relativity is made of two terms,
\begin{equation}\label{S}
S = \frac{c^3}{16\pi G}\int \ud^4x \,\sqrt{-g}\,R + S_\mathrm{matter}\left[\Psi, g\lab\right]\,.
\end{equation}
The first term represents the kinetic Einstein-Hilbert action for gravity and tells that the gravitational field $g\lab$ propagates like a pure spin-2 field. Here $R$ is the Ricci scalar and $g=\mathrm{det}(g\lab)$ is the determinant of the metric. The second term expresses the fact that all matter fields (collectively denoted by $\Psi$) are minimally coupled to the metric $g\lab$ which defines the physical lengths and times as measured in local laboratory experiments. The field equations are obtained by varying the action with respect to the metric (such that $\delta g\lab=0$ when $\vert x^\mu\vert \rightarrow \infty$) and form a system of ten differential equations of second order,
\begin{equation}\label{EE}
  G\ab[g,\partial g,\partial^2g] =
  \frac{8\pi G}{c^4} T\ab[g]\,.
\end{equation}
The Einstein tensor $G\ab\equiv R\ab-\frac{1}{2}R \, g\ab$ is generated by the matter stress-energy tensor $T\ab\equiv (2/\sqrt{-g})(\delta S_\mathrm{matter}/\delta g\lab)$. Four equations give, \textit{via} the contracted Bianchi identity, the conservation equation for the matter system as
\begin{equation}\label{EOM}
  \nabla_\mu G^{\alpha\mu}\equiv 0
  \quad \Longrightarrow \quad
  \nabla_\mu T^{\alpha\mu}=0\,,
\end{equation}
which must be solved conjointly with the Einstein field equations for the gravitational field. The matter equation \eqref{EOM} reads also
\begin{equation}\label{EOM'}
\partial_\mu\left(\sqrt{-g}\,T^\mu_\alpha\right) = \frac{1}{2}\sqrt{-g}\,T^{\mu\nu}\partial_\alpha g_{\mu\nu}\,.
\end{equation}

Let us introduce an asymptotically Minkowskian coordinate system such that the gravitational-wave amplitude, defined by $h\ab\equiv\sqrt{-g}\, g\ab - \eta\ab$,\footnote{Here, $g\ab$ denotes the contravariant metric, inverse of the covariant metric $g\lab$, and $\eta\ab$ represents an auxiliary Minkowski metric. We assume that our spatial coordinates are Cartesian so that $(\eta\ab)=(\eta\lab)=\mathrm{diag}(-1,1,1,1)$.} is divergenceless, \textit{i.e.} satisfies the de Donder or harmonic gauge condition 
\begin{equation}\label{dh}
\partial_\mu h^{\alpha\mu}= 0\,.
\end{equation}
With this coordinate choice the Einstein field equations can be recast into the d'Alembertian equation
\begin{equation}\label{Dalembert}
\Box h\ab = \frac{16\pi G}{c^4} \tau\ab\,.
\end{equation}
Here $\Box = \eta^{\mu\nu}\partial_\mu\partial_\nu$ is the usual (\textit{flat}-spacetime) d'Alembertian operator. The source term $\tau\ab$ can rightly be interpreted as the ``effective'' stress-energy distribution of the matter and gravitational fields in harmonic coordinates; note that $\tau\ab$ is not a tensor and we shall often call it a pseudo-tensor. It is conserved in the sense that
\begin{equation}\label{dtau}
\partial_\mu\tau^{\alpha\mu}= 0\,.
\end{equation}
This is equivalent to the condition of harmonic coordinates \eqref{dh} and to the covariant conservation \eqref{EOM} of the matter tensor. The pseudo-tensor is made of the contribution of matter fields described by $T\ab$, and of the gravitational contribution $\Lambda\ab$ which is a complicated functional of the gravitational field variable $h^{\mu\nu}$ and its first and second derivatives. Thus,
\begin{equation}\label{tauexpr}
\tau\ab = |g| T\ab+ \frac{c^4}{16\pi
G}\,\Lambda\ab[h,\partial h,\partial^2h]\,.
\end{equation}
The point is that $\Lambda\ab$ is at least quadratic in the field strength $h^{\mu\nu}$, so the field equations \eqref{Dalembert} are naturally amenable to a perturbative non-linear treatment. The general expression is
\begin{align}\label{Lambda}
  \Lambda^{\alpha\beta} =& - h^{\mu\nu} \partial_\mu\partial_\nu
  h^{\alpha\beta}+\partial_\mu h^{\alpha\nu} \partial_\nu h^{\beta\mu}
  +\frac{1}{2}g^{\alpha\beta}g_{\mu\nu}\partial_\rho h^{\mu\sigma}
  \partial_\sigma h^{\nu\rho} \nonumber \\
  & - g^{\alpha\mu}g_{\nu\sigma}\partial_\rho h^{\beta\sigma} \partial_\mu
  h^{\nu\rho} -g^{\beta\mu}g_{\nu\sigma}\partial_\rho h^{\alpha\sigma}
  \partial_\mu h^{\nu\rho} +g_{\mu\nu}g^{\rho\sigma}\partial_\rho
  h^{\alpha\mu} \partial_\sigma h^{\beta\nu} \nonumber \\
  & + \frac{1}{4}\bigl(2g^{\alpha\mu}g^{\beta\nu}-g^{\alpha\beta}g^{\mu\nu}\bigr)\Bigl(g_{\rho\sigma}g_{\epsilon\pi}-\frac{1}{2}
  g_{\sigma\epsilon}g_{\rho\pi}\Bigr) \partial_\mu h^{\rho\pi}
  \partial_\nu h^{\sigma\epsilon}\,.
\end{align}

To select a physically sensible solution of the field equations in the case of a bounded matter system, we must impose a boundary condition at infinity, namely the famous no-incoming radiation condition, which ensures that the system is truly isolated from other bodies in the Universe. In principle the no-incoming radiation condition is to be formulated at past null infinity $\mathcal{I}^-$. Here, we can simplify the formulation by taking advantage of the presence of the Minkowski background $\eta\lab$ to define the no-incoming radiation condition with respect to Minkowskian past null infinity say $\mathcal{I}_\eta^-$. Within approximate methods this is legitimate as we can view the gravitational field as propagating on the flat background $\eta\lab$; indeed $\eta\lab$ does exist at any finite order of approximation.

The no-incoming radiation condition should be such that it suppresses at $\mathcal{I}_\eta^-$ any homogeneous (regular in $\mathbb{R}^4$) solution of the d'Alembertian equation $\Box h_\mathrm{hom}=0$ in a neighborhood of $\mathcal{I}_\eta^-$. We have at our disposal the Kirchhoff formula which expresses $h_\mathrm{hom}$ at some field point $({\bf x}',t')$ in terms of its values and its derivatives on a sphere centered on ${\bf x}'$ with radius $\rho\equiv \vert\mathbf{x}'-\mathbf{x}\vert$ and at retarded time $t\equiv t'-\rho/c$,
\begin{equation}\label{Kirchhoff}
h_\mathrm{hom}(\mathbf{x}',t')=\int {\ud\Omega\over 4\pi}\biggl[{\partial\over
\partial \rho}(\rho h_\mathrm{hom}) +{1\over c}{\partial\over\partial t}(\rho
h_\mathrm{hom})\biggr]({\bf x},t)\,,
\end{equation}
where $\ud\Omega$ is the solid angle spanned by the unit direction $\mathbf{N}\equiv(\mathbf{x}-\mathbf{x}')/\rho$. From this formula we deduce the no-incoming radiation condition as the following limit at $\mathcal{I}_\eta^-$, \textit{i.e.} when $r\equiv\vert\mathbf{x}\vert\rightarrow +\infty$ with $t+r/c=$ const,\footnote{In fact we obtain also the auxiliary condition that $r\partial_\mu h\ab$ should be bounded near $\mathcal{I}_\eta^-$. This comes from the fact that $\rho$ differs from $r$ and we have $\rho = r-\mathbf{x}'\cdot\mathbf{n}+\mathcal{O}(1/r)$ with $\mathbf{n}=\mathbf{x}/r$.}
\begin{equation}\label{noincoming}
\lim_{\mathcal{I}_\eta^-}
\biggl[{\partial\over \partial r}(rh\ab)
+{1\over c}{\partial\over\partial t}(rh\ab)\biggr]=0\,.
\end{equation}
Now if $h\ab$ satisfies \eqref{noincoming}, so does the pseudo-tensor $\tau\ab$ built on it, and then it is clear that the retarded integral of $\tau\ab$ does satisfy the same condition. Thus we infer that the unique solution of the Einstein equation \eqref{Dalembert} reads 
\begin{equation}\label{integrodiff}
 h\ab = {16\pi G\over c^4}\,\Box ^{-1}_\mathrm{R} \tau\ab\,,
\end{equation}
where the retarded integral takes the standard form
\begin{equation}\label{Box}
(\Box ^{-1}_\mathrm{R} \tau\ab)({\bf x},t)\equiv-{1\over 4\pi}\int {\ud^3{\bf
x}'\over \vert\mathbf{x}-\mathbf{x}'\vert}\tau\ab\left(\mathbf{x}',t-\vert\mathbf{x}-\mathbf{x}'\vert/c\right)\,.
\end{equation}
Notice that since $\tau\ab$ depends on $h^{\mu\nu}$ and its derivatives, the equation \eqref{integrodiff} is to be viewed as an integro-differential equation equivalent to the Einstein equation \eqref{Dalembert} with no-incoming radiation condition.

\section{Post-Newtonian iteration in the near zone}
\label{sec3}

In this Section we proceed with the post-Newtonian iteration of the field equations in harmonic coordinates in the near zone of an isolated matter distribution. We have in mind a general hydrodynamical fluid, whose stress-energy tensor is smooth, \textit{i.e.} $T\ab\in C^\infty(\mathbb{R}^4)$. Thus the scheme \textit{a priori} excludes the presence of singularities; these will be dealt with in later Sections. 

Let us remind that the post-Newtonian approximation in ``standard'' form (\textit{e.g.} \cite{AD75, K80a, K80b}) is plagued with some apparently inherent difficulties, which crop up at some high post-Newtonian order like 3PN. Up to the 2.5PN order the approximation can be worked out without problems, and at the 3PN order the problems can generally be solved for each case at hand; but the problems worsen at higher orders. Historically these difficulties, even appearing at higher approximations, have cast a doubt on the actual soundness, from a theoretical point of view, of the post-Newtonian expansion. Practically speaking, they posed the question of the reliability of the approximation, when comparing the theory's predictions with very precise experimental results. In this Section and the next one we assess the nature of these difficulties --- are they purely technical or linked with some fundamental drawback of the approximation scheme? --- and eventually resolve them. 

We first distinguish the problem of \textit{divergences} in the post-Newtonian expansion: in higher approximations some divergent Poisson-type integrals appear. Recall that the post-Newtonian expansion replaces the resolution of an hyperbolic-like d'Alembertian equation by a perturbatively equivalent hierarchy of elliptic-like Poisson equations. Rapidly it is found during the post-Newtonian iteration that the right-hand-sides of the Poisson equations acquire a non-compact support (it is distributed all over space), and that the standard Poisson integral diverges because of the bound of the integral at spatial infinity, \textit{i.e.} $r\equiv \vert\mathbf{x}\vert\rightarrow +\infty$, with $t=$ const.

The divergencies are linked to the fact that the post-Newtonian expansion is actually a singular perturbation, in the sense that the coefficients of the successive powers of $1/c$ are not uniformly valid in space, typically blowing up at spatial infinity like some positive powers of $r$. We know for instance that the post-Newtonian expansion cannot be ``asymptotically flat'' starting at the 2PN or 3PN level, depending on the adopted coordinate system \cite{Rend92}. The result is that the standard Poisson integrals are in general badly-behaving at infinity. Trying to solve the post-Newtonian equations by means of the Poisson integral does not \textit{a priori} make sense. This does not mean that there are no solutions to the problem, but simply that the Poisson integral does not constitute the good solution of the Poisson equation in the context of post-Newtonian expansions. So the difficulty is purely of a technical nature, and will be solved once we succeed in finding the appropriate solution to the Poisson equation. 

We shall now prove (following \cite{PB02}) that the post-Newtonian expansion can be \textit{indefinitely} iterated without divergences.\footnote{An alternative solution to the problem of divergencies, proposed in \cite{FS83, F83}, is based on an initial-value formalism, which avoids the appearance of divergencies because of the finiteness of the integration region.} Let us denote by means of an overline the formal (infinite) post-Newtonian expansion of the field inside the source's near-zone, which is of the form
\begin{equation}\label{hbar}
 \overline{h}\ab({\mathbf
 x},t,c)=\sum_{n=2}^{+\infty}\,\frac{1}{c^n}\,\mathop{\overline{h}}_{n}{}\ab({\mathbf
 x},t,\ln c)\,. 
\end{equation}
The $n$-th post-Newtonian coefficient is naturally the factor of the $n$-th power of $1/c$; however, we know \cite{BD86} that the post-Newtonian expansion also involves some logarithms of $c$, included for convenience here into the definition of the coefficients $\overline{h}_{n}$. For the stress-energy pseudo-tensor \eqref{tauexpr} we have the same type of expansion,
\begin{equation}\label{taubar}
 \overline{\tau}\ab({\mathbf
 x},t,c)=\sum_{n=-2}^{+\infty}\,\frac{1}{c^n}\,\mathop{\overline{\tau}}_{n}{}\ab({\mathbf
 x},t,\ln c)\,. 
\end{equation}
The expansion starts with a term of order $c^2$ corresponding to the rest mass-energy ($\tau\ab$ has the dimension of an energy density). Here we shall always understand the infinite sums such as \eqref{hbar}--\eqref{taubar} in the sense of \textit{formal} series, \textit{i.e.} merely as an ordered collection of coefficients. Because of our consideration of regular extended matter distributions the post-Newtonian coefficients are smooth functions of space-time. 

Inserting the post-Newtonian ansatz into the Einstein field equation \eqref{Dalembert} and equating together the powers of $1/c$ results is an infinite set of Poisson-type equations ($\forall n\geq 2$),
\begin{equation}\label{Poisson}
\Delta\mathop{\overline{h}}_{n}{}\ab =16\pi G \mathop{\overline{\tau}}_{n-4}\!\!{}\ab+\partial_{t}^{2}\!\!\mathop{\overline{h}}_{n-2}\!\!{}\ab
\,. \end{equation}
The second term comes from the split of the d'Alembertian into a Laplacian and a second time derivative: $\Box=\Delta-\frac{1}{c^2}\partial_t^2$. This term is zero when $n=2$ and $3$. We proceed by induction, \textit{i.e.} fix some post-Newtonian order $n$, assume that we succeeded in constructing the sequence of previous coefficients ${}_p\overline{h}$ for $p\leq n-1$, and from this show how to infer the next-order coefficient ${}_n\overline{h}$.

To cure the problem of divergencies we introduce a generalized solution of the Poisson equation with non-compact support source, in the form of an appropriate \textit{finite part} of the usual Poisson integral obtained by regularization of the bound at infinity by means of a specific process of analytic continuation. For any source term like ${}_n\overline{\tau}$, we multiply it by the ``regularization'' factor
\begin{equation}\label{rB}
\vert\widetilde{\mathbf{x}}\vert^B\equiv\left|\frac{\mathbf{x}}{r_0}\right|^B\,,
\end{equation}
where $B\in\C$ is a complex number and $r_0$ denotes an arbitrary length scale. Only then do we apply the Poisson integral, which therefore defines a certain function of $B$. The well-definedness of that integral heavily relies on the behavior of the integrand at the bound at infinity. There is no problem with the vicinity of the origin inside the source because of the smoothness of the pseudo-tensor. Then one can prove \cite{PB02} that the latter function of $B$ generates a (unique) analytic continuation down to a neighborhood of the value of interest $B=0$, except at $B=0$ itself, at which value it admits a Laurent expansion with multiple poles up to some finite order. Then, we consider the Laurent expansion of that function when $B\rightarrow 0$ and pick up the finite part, or coefficient of the zero-th power of $B$, of that expansion. This defines our generalized Poisson integral:
\begin{equation}\label{genpoiss}
\Delta^{-1}\big[\,\mathop{\overline{\tau}}_{n}{}\ab\big](\mathbf{x},t) \equiv
-\frac{1}{4\pi}\,\FP\,\int \,\frac{\ud^3\mathbf{x}'}{\vert\mathbf{x}-\mathbf{x}'\vert}\,\vert\widetilde{\mathbf{x}}'\vert^B\,\mathop{\overline{\tau}}_{n}{}\ab(\mathbf{x}',t) \,. 
\end{equation} 
The integral extends over all three-dimensional space but with the latter finite-part regularization at infinity denoted $\FP$. The main properties of our generalized Poisson operator is that it does solve the Poisson equation, namely
\begin{equation}\label{checkpoiss}
\Delta\left(\Delta^{-1}\big[\,\mathop{\overline{\tau}}_{n}{}\ab\big]\right) =
\mathop{\overline{\tau}}_{n}{}\ab \,, 
\end{equation} 
and that the so defined solution $\Delta^{-1}{}_n\overline{\tau}$ owns the same properties as its source ${}_n\overline{\tau}$, \textit{i.e.} the smoothness and the same type of behavior at infinity.

The most general solution of the Poisson equation \eqref{Poisson} will be obtained by application of the previous generalized Poisson operator to the right-hand-side of \eqref{Poisson}, and augmented by the most general \textit{homogeneous} solution of the Poisson equation. Thus, we can write
\begin{equation}\label{hngen}
\mathop{\overline{h}}_{n}{}\ab =16\pi
G\,\Delta^{-1}\big[\mathop{\overline{\tau}}_{n-4}\!\!{}\ab\big]
+\partial_{t}^{2}\,\Delta^{-1}\big[\mathop{\overline{h}}_{n-2}\!\!{}\ab\big] + \sum_{\ell=0}^{+\infty}\mathop{B}_{n}{}_L\ab\!(t)\,{\hat
x}_L\,. 
\end{equation}
The last term represents the general solution of the Laplace equation which is regular at the origin $r\equiv\vert\mathbf{x}\vert=0$. It can be written, using the symmetric-trace-free (STF) language, as a multipolar series of terms of the type $\hat{x}_L$,\footnote{Here $L=i_1\cdots i_\ell$ denotes a multi-index composed of $\ell$ multipolar spatial indices $i_1, \cdots, i_\ell$ (ranging from 1 to 3); $x_L \equiv x_{i_1}\cdots x_{i_\ell}$ is the product of $\ell$ spatial vectors $x_i$; $\partial_L = \partial_{i_1}\cdots \partial_{i_\ell}$ is the product of $\ell$ partial derivatives $\partial_i=\partial/\partial x^i$; in the case of summed-up (dummy) multi-indices $L$, we do not write the $\ell$ summations from 1 to 3 over their indices; the STF projection is indicated with a hat, \textit{i.e.} $\hat{x}_L\equiv\mathrm{STF}[x_L]$ and similarly $\hat{\partial}_L\equiv\mathrm{STF}[\partial_L]$, or sometimes using brackets surrounding the indices, $x_{<L>}\equiv\hat{x}_L$.} and multiplied by some STF-tensorial functions of time ${}_n B_L(t)$. These functions will be associated with the radiation reaction of the field onto the source; they will depend on which boundary conditions are to be imposed on the gravitational field at infinity from the source.

It is now trivial to iterate the process. We substitute for ${}_{n-2}\overline{h}$ in the right-hand-side of \eqref{hngen} the same expression but with $n$ replaced by $n-2$, and similarly come down until we stop at either one of the coefficients ${}_{0}\overline{h}=0$ or ${}_{1}\overline{h}=0$. At this point ${}_{n}\overline{h}$ is expressed in terms of the ``previous'' ${}_{p}\overline{\tau}$'s and ${}_{p}B_L$'s with $p\leq n-2$. To finalize the process we introduce what we call the operator of the ``instantaneous'' potentials $\Box^{-1}_\mathrm{I}$. Our notation is chosen to contrast with the standard operators of the retarded and advanced potentials $\Box^{-1}_\mathrm{R}$ and $\Box^{-1}_\mathrm{A}$, see \eqref{Box}. However, beware of the fact that unlike $\Box^{-1}_{\mathrm{R},\mathrm{A}}$ the operator $\Box^{-1}_\mathrm{I}$ will be defined only when acting on a post-Newtonian series such as $\overline{\tau}$. Indeed, we pose
\begin{equation}\label{BoxI}
\Box^{-1}_\mathrm{I}\big[\,\overline{\tau}\ab\big] \equiv
\sum_{k=0}^{+\infty}\left(\frac{\partial}{c\partial
t}\right)^{\!\!2k}\Delta^{-k-1}\big[\,\overline{\tau}\ab\big]\,, \end{equation} 
where $\Delta^{-k-1}$ is the $k$-th iteration of the operator \eqref{genpoiss}. It is readily checked that in this way we have a solution of the source-free d'Alembertian equation,
\begin{equation}\label{BoxBoxI}
\Box\left(\Box^{-1}_\mathrm{I}\big[\,\overline{\tau}\ab\big]\right) =
\overline{\tau}\ab \,.
\end{equation} 
On the other hand, the homogeneous solution in \eqref{hngen} will yield by iteration an homogeneous solution of the d'Alembertian equation which is necessarily regular at the origin. Hence it should be of the \textit{anti-symmetric} type, \textit{i.e.} be made of the difference between a retarded multipolar wave and the corresponding advanced wave. We shall therefore introduce a new definition for some STF-tensorial functions $A_L(t)$ parametrizing those advanced-minus-retarded free waves. It will not be difficult to relate the post-Newtonian expansion of $A_L(t)$ to the functions ${}_n B_L(t)$ which were introduced in \eqref{hngen}. Finally the most general post-Newtonian solution, iterated \textit{ad infinitum} and without any divergences, is obtained into the form
\begin{equation}\label{hgen1}
\overline{h}\ab=\frac{16\pi G}{c^4} \,\Box^{-1}_\mathrm{I}\big[\,\overline{\tau}\ab\big]
- \frac{4G}{c^4}\sum^{+\infty}_{\ell=0} \frac{(-)^\ell}{\ell!}\hat{\partial}_L \left\{ \frac{A\ab_L
(t-r/c)-A\ab_L (t+r/c)}{2r} \right\}\,. 
\end{equation} 
We shall refer to the $A_L(t)$'s as the \textit{radiation-reaction} functions. If we stay at the level of the post-Newtonian iteration which is confined into the near zone we cannot do more than \eqref{hgen1}; there is no means to compute the radiation-reaction functions $A_L(t)$. We are here touching the second problem faced by the standard post-Newtonian approximation.

\section{Post-Newtonian expansion calculated by matching}
\label{sec4}
 
As we now understand this problem is that of the limitation to the near zone. Indeed the post-Newtonian expansion assumes that all retardations $r/c$ are small, so it can be viewed as a formal \textit{near-zone} expansion when $r\rightarrow 0$, valid only in the region surrounding the source that is of small extent with respect to the wavelength of the emitted radiation: $r\ll \lambda$. As we have seen, a consequence is that the post-Newtonian coefficients blow up at infinity, when $r\rightarrow +\infty$. It is thus not possible, \textit{a priori}, to implement within the post-Newtonian scheme the physical information that the matter system is isolated from the rest of the Universe. The no-incoming radiation condition imposed at past null infinity $\mathcal{I}_\eta^-$ cannot be taken into account, \textit{a priori}, within the scheme.

The near-zone limitation can be circumvented to the lowest post-Newtonian orders by considering \textit{retarded} integrals that are formally expanded when $c\to +\infty$ as series of ``instantaneous'' Poisson-like integrals \cite{AD75}. This procedure works well up to the 2.5PN level and has been shown to correctly fix the dominant radiation reaction term at the 2.5PN order \cite{K80a, K80b}. Unfortunately such a procedure assumes fundamentally that the gravitational field, after expansion of all retardations $r/c\to 0$, depends on the state of the source at a single time $t$, in keeping with the instantaneous character of the Newtonian interaction. However, we know that the post-Newtonian field (as well as the source's dynamics) will cease at some stage to be given by a functional of the source parameters at a single time, because of the imprint of gravitational-wave tails in the near zone field, in the form of some modification of the radiation reaction force at the 1.5PN relative order \cite{BD88, B93}. Since the reaction force is itself of order 2.5PN this means that the formal post-Newtonian expansion of retarded Green functions is no longer valid starting at the 4PN order. 

The solution of the problem resides in the matching of the near-zone field to the exterior field, a solution of the vacuum equations outside the source which has been developed in previous works using some post-\textit{Minkowskian} and multipolar expansions \cite{BD86, BD92}. In the case of post-Newtonian sources, the near zone, \textit{i.e.} $r\ll\lambda$, covers entirely the source, because the source's radius itself is such that $a\ll\lambda$. Thus the near zone overlaps with the exterior zone where the multipole expansion is valid. Matching together the post-Newtonian and multipolar-post-Minkowskian solutions in this overlapping region is an application of the method of matched asymptotic expansions, which has frequently been applied in the present context, both for radiation-reaction \cite{BuTh70, Bu71, BD88, B93} and wave-generation \cite{BD89, DI91a, B98mult} problems.

In the previous Section we obtained the most general solution \eqref{hgen1} for the post-Newtonian expansion, as parametrized by the set of unknown radiation-reaction functions $A_L(t)$. We shall now impose the \textit{matching} condition
\begin{equation}\label{matching}
\mathcal{M}(\overline{h}\ab)\equiv\overline{\mathcal{M}(h\ab)}\,,
\end{equation} 
telling that the \textit{multipole} decomposition of the post-Newtonian expansion $\overline{h}$ of the inner field, agrees with the \textit{near-zone} expansion of the multipole expansion $\mathcal{M}(h)$ of the external field. Here the calligraphic letter $\mathcal{M}$ stands for the multipole decomposition or far-zone expansion, while the overbar denotes the post-Newtonian or near-zone expansion. The matching equation results from the numerical equality $\overline{h}=\mathcal{M}(h)$, clearly verified in the exterior part of the  near-zone, namely our overlapping region $a<r\ll\lambda$. The left-hand-side is expanded when $r\rightarrow +\infty$ yielding $\mathcal{M}(\overline{h})$ while the right-hand-side is expanded when $r\rightarrow 0$ leading to $\overline{\mathcal{M}(h)}$. The matching equation is thus physically justified only for post-Newtonian sources, for which the exterior near-zone exists. It is actually a \textit{functional} identity; it identifies, \textit{term-by-term}, two asymptotic expansions, each of them being formally taken outside its own domain of validity. In the present context, the matching equation insists that the infinite \textit{far-zone} expansion ($r\rightarrow \infty$) of the inner post-Newtonian field is identical to the infinite \textit{near-zone} expansion ($r\rightarrow 0$) of the exterior multipolar field. Let us now state that \eqref{matching}, plus the condition of no-incoming radiation, permits determining all the unknowns of the problem: \textit{i.e.}, at once, the external multipolar decomposition $\mathcal{M}(h)$ \textit{and} the radiation-reaction functions $A_L$ and hence the inner post-Newtonian expansion $\overline{h}$.

When applied to a multipole expansion such as that of the pseudo-tensor, \textit{i.e.} $\mathcal{M}(\tau\ab)$, we have to define a special type of generalized inverse d'Alembertian operator, built on the standard retarded integral \eqref{Box}, \textit{viz}
\begin{equation}\label{retarded}
\Box^{-1}_\mathrm{R}\big[\mathcal{M}(\tau\ab)\big](\mathbf{x},t) \equiv -\frac{1}{4\pi}\,\FP\,\int\,\frac{\ud^3\mathbf{x}'}{\vert\mathbf{x}-\mathbf{x}'\vert}\,\vert\widetilde{\mathbf{x}}'\vert^B\mathcal{M}(\tau\ab)(\mathbf{x}',
t-\vert\mathbf{x}-\mathbf{x}'\vert/c)\,. 
\end{equation}
Like \eqref{Box} this integral extends over the whole three-dimensional space, but a regularization factor $\vert\widetilde{\mathbf{x}}'\vert^B$ given by \eqref{rB} has been ``artificially'' introduced for application of the finite part operation $\FP$. The reason for introducing such regularization is to cure the divergencies of the integral when $\vert\mathbf{x}'\vert\rightarrow 0$; these are coming from the fact that the multipolar expansion is singular at the origin. We notice that this regularization factor is the same as the one entering the generalized Poisson integral \eqref{genpoiss}, however its role is different, as it takes care of the bound at $\vert\mathbf{x}'\vert = 0$ rather than at infinity. We easily find that this new object is a particular retarded solution of the wave equation, 
\begin{equation}\label{BoxM}
\Box\left(\Box_\mathrm{R}^{-1}\big[\,\mathcal{M}(\tau\ab)\big]\right) = \mathcal{M}(\tau\ab) \,.
\end{equation} 
Therefore $\mathcal{M}(h\ab)$ should be given by that solution plus a \textit{retarded} homogeneous solution of the d'Alembertian equation (imposing the no-incoming radiation condition), \textit{i.e.} be of the type
\begin{equation}\label{Multh}
\mathcal{M}(h\ab) = \frac{16\pi G}{c^4}\,\Box^{-1}_\mathrm{R}\big[\mathcal{M}(\tau\ab)\big] - \frac{4G}{c^4}\sum^{+\infty}_{\ell=0} \frac{(-)^\ell}{\ell!}\hat{\partial}_L \left\{ \frac{F\ab_L
(t-r/c)}{r} \right\}\,.
\end{equation}
Now the matching equation \eqref{matching} will determine both the $A_L$'s in \eqref{hgen1} and the $F_L$'s in \eqref{Multh}. We summarize the results which have been obtained in \cite{PB02}.
 
The functions $F_L(t)$ will play an important role in the following, because they appear as the multipole moments of a general post-Newtonian source as seen from its exterior near zone. Their closed-form expression obtained by matching reads
\begin{equation} \label{FL}
F\ab_L(t) =
\FP\,\int\,\ud^3\mathbf{x}\,\vert\widetilde{\mathbf{x}}\vert^B\,\hat{x}_L\int_{-1}^{+1}\ud z\,
\delta_\ell(z)\,\overline{\tau}\ab\,\left(\mathbf{x},t
-z\vert\mathbf{x}\vert/c\right) \,.
\end{equation}
Again the integral extends over all space but the bound at infinity (where the post-Newtonian expansion becomes singular) is regularized by means of the same finite part. The $z$-integration involves a weighting function $\delta_\ell(z)$ defined by
\begin{equation} \label{deltal}
\delta_\ell (z) = \frac{(2\ell+1)!!}{2^{\ell+1} \ell!} \,(1-z^2)^\ell\,.
\end{equation}
The integral of that function is normalized to one: $\int_{-1}^{+1}\ud z\,\delta_\ell (z) = 1$. Furthermore it approaches the Dirac function in the limit of large multipoles: $\lim_{\ell\rightarrow\infty}\delta_\ell (z)=\delta (z)$. The multipole moments \eqref{FL} are physically valid only for post-Newtonian sources. As such, they must be considered only in a perturbative post-Newtonian sense. With the result \eqref{FL} the multipole expansion \eqref{Multh} is fully determined and will be exploited in the next Section. 

Concerning the near-zone field \eqref{hgen1} we find that the radiation-reaction functions $A_L$ are composed of the multipole moments $F_L$ which will also characterize ``linear-order'' radiation reaction effects starting at 2.5PN order, and of an extra contribution $R_L$ which will be due to non-linear effets in the radiation reaction and turn out to arise at 4PN order. Thus,
\begin{equation} \label{AFR}
A\ab_L=F\ab_L+R\ab_L\,,
\end{equation}
where $F_L$ is given by \eqref{FL} and where $R_L$ is defined from the multipole expansion of the pseudo-tensor as
\begin{equation}\label{RL}
R\ab_L(t) =
\FP\,\int\,\ud^3\mathbf{x}\,\vert\widetilde{\mathbf{x}}\vert^B\,\hat{x}_L\int_{1}^{+\infty}\ud z\,
\gamma_\ell(z)\,\mathcal{M}(\tau\ab)\,\left(\mathbf{x},t
-z\vert\mathbf{x}\vert/c\right) \,.
\end{equation}
Here the regularization deals with the bound of the integral at $\vert\mathbf{x}\vert=0$. Since the variable $z$ extends up to infinity these functions truly depend on the whole past-history of the source. The weighting function therein is simply given by $\gamma_\ell(z)\equiv-2\delta_\ell(z)$, this definition being motivated by the fact that the integral of that function is normalized to one: $\int_1^{+\infty} \ud z\,\gamma_\ell(z) = 1$.\footnote{This integral is \textit{a priori} divergent, however its value can be obtained by invoking complex analytic continuation in $\ell\in\C$.} The specific contributions due to $R_L$ in the post-Newtonian metric \eqref{hgen1} are associated with tails of waves \cite{BD88, B93}. The fact that the external multipolar expansion $\mathcal{M}(\tau)$ is the source term for the function $R_L$, and therefore will enter the expression of the near-zone metric \eqref{hgen1}, is a result of the matching condition \eqref{matching} and reflects of course the no-incoming radiation condition imposed at $\mathcal{I}_\eta^-$.

The post-Newtonian metric \eqref{hgen1} is now fully determined. However let us now derive an interesting alternative formulation of it \cite{BFN05}. To this end we introduce still another object which will be made of the expansion of the standard retarded integral \eqref{Box} when $c\rightarrow\infty$, but acting on a post-Newtonian source term $\overline{\tau}$, 
\begin{equation} \label{BoxR}
\Box_\mathrm{R}^{-1}\big[\,\overline{\tau}\ab\big](\mathbf{x},t) \equiv
-\frac{1}{4\pi}\sum_{n=0}^{+\infty}\frac{(-)^n}{n!}\left(\frac{\partial}{c\,\partial
t}\right)^{\!n}\,\FP\,\int
\ud^3\mathbf{x}'\,\vert\widetilde{\mathbf{x}}'\vert^B\,\vert\mathbf{x}
-\mathbf{x}'\vert^{n-1}\,\overline{\tau}\ab(\mathbf{x}',t) \,.
\end{equation}
Each of the terms is regularized by means of the finite part to deal with the bound at infinity where the post-Newtonian expansion is singular. This regularization is crucial and the object should carefully be distinguished from the ``global'' solution $\Box_\mathrm{R}^{-1}[\tau]$ defined by \eqref{integrodiff} and in which the pseudo-tensor is \textit{not} expanded in post-Newtonian fashion. We emphasize that \eqref{BoxR} constitutes merely the \textit{definition} of a (formal) post-Newtonian expansion, each term of which being built from the post-Newtonian expansion of the pseudo-tensor. Such a definition is of interest because it corresponds to what one would intuitively think as the ``natural'' way of performing the post-Newtonian iteration, \textit{i.e.} by Taylor expanding the retardations as in \cite{AD75}. Moreover, each of the terms of the series \eqref{BoxR} is mathematically well-defined thanks to the finite part, and can therefore be implemented in practical computations. The point is that \eqref{BoxR} solves, in a post-Newtonian sense, the wave equation,
\begin{equation}\label{BoxBoxR}
\Box\left(\Box_\mathrm{R}^{-1}\big[\,\overline{\tau}\ab\big]\right) =
\overline{\tau}\ab \,, 
\end{equation} 
so constitutes a good prescription for a particular solution of the wave equation --- as legitimate a prescription as \eqref{BoxI}. Therefore \eqref{BoxI} and \eqref{BoxR} should differ by an homogeneous solution of the wave equation which is necessarily of the anti-symmetric type. Detailed investigations yield
\begin{equation}\label{BoxRI}
\Box_\mathrm{R}^{-1}\big[\,\overline{\tau}\ab\big] = \Box_\mathrm{I}^{-1}\big[\,\overline{\tau}\ab\big]
- \frac{1}{4\pi}\sum^{+\infty}_{\ell=0} \frac{(-)^\ell}{\ell!}\hat{\partial}_L \left\{ \frac{F\ab_L
(t-r/c)-F\ab_L (t+r/c)}{2r} \right\}\,,
\end{equation} 
in which the homogeneous solution is parametrized precisely by the multipole-moment functions $F_L(t)$. This formula is the basis of our writing of the new form of the post-Newtonian expansion. Indeed, by combining \eqref{hgen1} and \eqref{BoxRI}, we nicely get
\begin{equation}\label{hgen2}
\overline{h}\ab=\frac{16\pi G}{c^4}\,\Box_\mathrm{R}^{-1}\big[\,\overline{\tau}\ab\big]
- \frac{4G}{c^4}\sum^{+\infty}_{\ell=0} \frac{(-)^\ell}{\ell!}\hat{\partial}_L \left\{ \frac{R\ab_L
(t-r/c)-R\ab_L (t+r/c)}{2r} \right\}\,,
\end{equation} 
which is our final expression for the general solution of the post-Newtonian field in the near-zone of any isolated matter distribution. This expression is probably the most convenient and fruitful when doing practical applications. 

We recognize in the first term of \eqref{hgen2} (notwithstanding the finite part therein) the old way of performing the post-Newtonian expansion as it was advocated by Anderson \& DeCanio \cite{AD75}. For computations limited to the 3.5PN order, \textit{i.e.} up to the level of the 1PN correction to the radiation reaction force, such first term is sufficient. However, at the 4PN order there is a fundamental breakdown of this scheme and it becomes necessary to take into account the second term in \eqref{hgen2} which corresponds to non-linear radiation reaction effects associated with tails.

Note that the post-Newtonian solution, in either form \eqref{hgen1} or \eqref{hgen2}, has been obtained without imposing the condition of harmonic coordinates in an explicit way, see \eqref{dh}. We have simply matched together the post-Newtonian and multipolar expansions, satisfying the ``relaxed'' Einstein field equations \eqref{Dalembert} in their respective domains, and found that the matching determines uniquely the solution. An important check (carried out in \cite{PB02, BFN05}) is therefore to verify that the harmonic coordinate condition \eqref{dh} is indeed satisfied as a consequence of the conservation of the pseudo-tensor \eqref{dtau}, so that we really grasp a solution of the full Einstein field equations. 

\section{Multipole moments of a post-Newtonian source}
\label{sec5}

The multipole expansion of the field outside a general post-Newtonian source has been obtained in the previous Section as\footnote{An alternative formulation of the multipole expansion for a post-Newtonian source, with non-STF multipole moments, has been developed by Will and collaborators \cite{WW96,PW00,PW02}.}
\begin{equation}\label{Mhab}
\mathcal{M}(h\ab) = - \frac{4G}{c^4}\sum^{+\infty}_{\ell=0} \frac{(-)^\ell}{\ell!}\hat{\partial}_L \left\{ \frac{F\ab_L
(t-r/c)}{r} \right\} + u\ab\,,
\end{equation}
where the multipole moments are explicitly given by \eqref{FL}, and the second piece reflects the non-linearities of the Einstein field equations and reads
\begin{equation}\label{uab}
u\ab = \Box^{-1}_\mathrm{R}\big[\mathcal{M}(\Lambda\ab)\big] \,.
\end{equation}
To write the latter expression we have used the fact that since the matter tensor $T\ab$ has a spatially compact support we have $\mathcal{M}(T\ab)=0$. Thus $u\ab$ is indeed generated by the non-linear gravitational source term \eqref{Lambda}. We notice that the divergence of this piece, say $w^\alpha\equiv\partial_\mu u^{\alpha\mu}$, is a retarded homogeneous solution of the wave equation, \textit{i.e.} of the same type as the first term in \eqref{Mhab}. Now from $w^\alpha$ we can construct a secondary object $v\ab$ which is also a retarded homogeneous solution of the wave equation, and furthermore whose divergence satisfies $\partial_\mu v^{\alpha\mu}=-w^\alpha$, so that it will cancel the divergence of $u\ab$ (see \cite{B98mult} for details). With the above construction of $v\ab$ we are able to define the following combination,
\begin{equation}\label{h1ab}
G\,h_{(1)}\ab \equiv - \frac{4G}{c^4}\sum^{+\infty}_{\ell=0} \frac{(-)^\ell}{\ell!}\hat{\partial}_L \left\{ \frac{F\ab_L
(t-r/c)}{r} \right\} - v\ab\,, 
\end{equation}
which will constitute the \textit{linearized} approximation to the multipolar expansion $\mathcal{M}(h\ab)$ outside the source. Then we have 
\begin{equation}\label{h1uv}
\mathcal{M}(h\ab)=G\,h_{(1)}\ab+u\ab+v\ab\,.
\end{equation}
Having singled out such linearized part, it is clear that the sum of the second and third terms should represent the non-linearities in the external field. If we index those non-linearities by Newton's constant $G$, then we can prove indeed that $u\ab+v\ab=\mathcal{O}(G^2)$. More precisely we can decompose $u\ab+v\ab$ as a complete non-linearity or ``post-Minkowskian'' expansion of the type
\begin{equation}\label{PM}
u\ab+v\ab = \sum_{m=2}^{+\infty}\,G^m\,h_{(m)}\ab\,.
\end{equation}
One can effectively define a post-Minkowskian ``algorithm'' \cite{BD86, B98mult} able to construct the non-linear series up to any post-Minkowskian order $m$. The post-Minkows-kian expansion represents the most general solution of the Einstein field equations in harmonic coordinates valid in the vacuum region outside an isolated source.
 
The above linearized approximation $h_{(1)}$ solves the linearized vacuum Einstein field equations in harmonic coordinates and it can be decomposed into multipole moments in a standard way \cite{Th80}. Modulo an infinitesimal gauge transformation preserving the harmonic gauge, namely
\begin{equation}\label{hk}
h_{(1)}\ab=k_{(1)}\ab+\partial^\alpha\varphi_{(1)}^\beta+\partial^\beta\varphi_{(1)}^\alpha-\eta\ab\partial_\mu\varphi_{(1)}^\mu\,,
\end{equation}
where the infinitesimal gauge vector $\varphi_{(1)}^\alpha$ satisfies $\Box\varphi_{(1)}^\alpha=0$, we can decompose\footnote{The supersript $(k)$ refers to $k$ time derivatives of the moments; $\varepsilon_{abc}$ is the
Levi-Civita antisymmetric symbol such that $\varepsilon_{123}=1$. From here on the spatial indices such as $i$, $j$, $\dots$ will be raised and lowered with the Kronecker metric $\delta_{ij}$. They will be located lower or upper depending on context.}
\begin{subequations}\label{kab}
\begin{align}
    k^{00}_{(1)} &= -\frac{4}{c^2}\sum_{\ell\geq 0}
      \frac{(-)^\ell}{\ell !} \partial_L \left( \frac{1}{r} \mathrm{I}_L
      \right)\,, \\ k^{0i}_{(1)} &= 
  \frac{4}{c^3}\sum_{\ell\geq 1} \frac{(-)^\ell}{\ell !} \left\{
      \partial_{L-1} \left( \frac{1}{r} \mathrm{I}_{iL-1}^{(1)}
      \right) + \frac{\ell}{\ell+1} \varepsilon_{iab} \partial_{aL-1}
      \left( \frac{1}{r} \mathrm{J}_{bL-1} \right)\right\}\,, \\ 
k^{ij}_{(1)} &= -\frac{4}{c^4}\sum_{\ell\geq 2}
      \frac{(-)^\ell}{\ell !} \left\{ \partial_{L-2} \left( \frac{1}{r}
      \mathrm{I}_{ijL-2}^{(2)} \right) + \frac{2\ell}{\ell+1}
      \partial_{aL-2} \left( \frac{1}{r} \varepsilon_{ab(i}
      \mathrm{J}_{j)bL-2}^{(1)} \right)\right\}\,.
  \end{align}
\end{subequations}
This decomposition defines two types of multipole moments, both assumed to be STF: the mass-type $\mathrm{I}_L(u)$ and the current-type $\mathrm{J}_L(u)$. These moments can be arbitrary functions of the retarded time $u\equiv t-r/c$, except that the monopole and dipole moments (having $\ell\leq 1$) satisfy standard conservation laws, namely
\begin{equation}\label{conserv}
\mathrm{I}^{(1)} = \mathrm{I}_i^{(2)} = \mathrm{J}_i^{(1)} = 0 \,.
\end{equation}
The gauge transformation vector admits a decomposion in similar fashion,
\begin{subequations}\label{phia}
\begin{align}
\varphi^0_{(1)} =& {4\over c^3}\sum_{\ell\geq 0} {(-)^\ell\over
    \ell !} \partial_L \left( {1\over r} \mathrm{W}_L \right), \\ 
\varphi^i_{(1)} =& -{4\over c^4}\sum_{\ell\geq 0}
    {(-)^\ell\over \ell !} \partial_{iL} \left( {1\over r} \mathrm{X}_L
    \right) \\  & -{4\over c^4}\sum_{\ell\geq 1}
    {(-)^\ell\over \ell !} \left\{ \partial_{L-1} \left( {1\over r}
    \mathrm{Y}_{iL-1} \right) + {\ell\over \ell+1} \varepsilon_{iab}
    \partial_{aL-1} \left( {1\over r} \mathrm{Z}_{bL-1}
    \right)\right\}\,.
\end{align}
\end{subequations}
The six sets of STF multipole moments $\mathrm{I}_L$, $\mathrm{J}_L$, $\mathrm{W}_L$, $\mathrm{X}_L$, $\mathrm{Y}_L$ and $\mathrm{Z}_L$ will collectively be called the multipole moments of the source. They contain the full physical information about any isolated source as seen from its exterior near zone. Actually it should be clear that the main moments are $\mathrm{I}_L$ and $\mathrm{J}_L$ because the other moments $\mathrm{W}_L$, $\cdots$, $\mathrm{Z}_L$ parametrize a linear gauge transformation and thus have no physical implications at the linearized order. However because the theory is covariant with respect to non-linear diffeomorphisms and not merely with respect to linear gauge transformations, the moments $\mathrm{W}_L$, $\cdots$, $\mathrm{Z}_L$ do play a physical role starting at the non-linear level. We shall occasionaly refer to the moments $\mathrm{W}_L$, $\mathrm{X}_L$, $\mathrm{Y}_L$ and $\mathrm{Z}_L$ as the gauge moments.

To express in the best way the source multipole moments, we introduce the following notation for combinations of components of the pseudo-tensor $\overline{\tau}\ab$,
\begin{subequations}\label{Sigma}
\begin{align}
 \overline{\Sigma} \equiv& {\overline{\tau}^{00} +\overline{\tau}^{ii}\over c^2}\,,\\
 \overline{\Sigma}_i \equiv& {\overline{\tau}^{0i}\over c}\,, \\
\overline{\Sigma}_{ij} \equiv& \overline{\tau}^{ij}\,, 
 \end{align}
\end{subequations}
where $\overline{\tau}^{ii} \equiv\delta_{ij}\overline\tau^{ij}$. Here the overbar reminds us that we are exclusively dealing with post-Newtonian-expanded expressions, \textit{i.e.} formal series of the type \eqref{taubar}. Then the general expressions of the ``main'' source multipole moments $\mathrm{I}_L$ and $\mathrm{J}_L$ in the case of the time-varying moments for which $\ell\geq 2$, are
\begin{subequations}\label{IJL}
\begin{align}
    \mathrm{I}_L(u) =& \FP \int \ud^3 \mathbf{x}\,\vert\widetilde{\mathbf{x}}\vert^B
      \int^1_{-1} \ud z\,\biggl\{ \delta_\ell\,\hat x_L\overline{\Sigma} -
      \frac{4(2\ell+1)}{c^2(\ell+1)(2\ell+3)} \delta_{\ell+1}\, \hat x_{iL}
      \overline{\Sigma}^{(1)}_i \nonumber\\ & \qquad \qquad \qquad \qquad 
      + \frac{2(2\ell+1)}{c^4(\ell+1)(\ell+2)(2\ell+5)} \delta_{\ell+2}\,
      \hat x_{ijL} \overline{\Sigma}^{(2)}_{ij} \biggr\} \,, \\ \mathrm{J}_L(u) =& \FP
      \int \ud^3 \mathbf{x}\,\vert\widetilde{\mathbf{x}}\vert^B\int^1_{-1} \ud z \, 
\varepsilon_{ab \langle i_\ell}
      \biggl\{ \delta_\ell\, {\hat x}_{L-1 \rangle a} \overline{\Sigma}_b
 \nonumber\\ 
     & \qquad \qquad \qquad \qquad -
      \frac{2\ell+1}{c^2(\ell+2)(2\ell+3)} \delta_{\ell+1}\, \hat x_{L-1 \rangle
 ac}
      \overline{\Sigma}^{(1)}_{bc} \biggr\}\,.
\end{align}
\end{subequations}
The integrands are computed at the spatial point $\mathbf{x}$ and at time $u+z \vert\mathbf{x}\vert/c$, where $u=t-r/c$ is the retarded time at which are evaluated the moments. We recall that $z$ is the argument of the function $\delta_\ell$ defined in \eqref{deltal}. Similarly we can write the expressions of the gauge-type moments $\mathrm{W}_L$, $\cdots$, $\mathrm{Z}_L$. Notice that the source multipole moments \eqref{IJL} have no invariant meaning; they are defined for the harmonic coordinate system we have chosen.

Of what use are these results for the multipole moments $\mathrm{I}_L$ and $\mathrm{J}_L$? From \eqref{kab} these moments parametrize the linearized metric $h_{(1)}$ which is the ``seed'' of an infinite post-Minkowskian algorithm symbolized by \eqref{PM}. For a specific application, \textit{i.e.} for a specific choice of matter tensor like the one we shall describe in Section \ref{sec8}, the expressions \eqref{IJL} have to be worked out up to a given post-Newtonian order. The moments should then be inserted into the post-Minkowskian series \eqref{PM} for the computation of the non-linearities. The result will be in the form of a non-linear multipole decomposition depending on the source moments $\mathrm{I}_L$, $\mathrm{J}_L$, $\cdots$, $\mathrm{Z}_L$, say 
\begin{equation}\label{MhabPM}
\mathcal{M}(h\ab)=\sum_{m=1}^{+\infty}\,G^m\,h_{(m)}\ab[\mathrm{I}_L,\mathrm{J}_L,\cdots]\,.
\end{equation}
In the next Section we shall expand this metric at (retarded) infinity from the source in order to obtain the observables of the gravitational radiation field. 

\section{Radiation field and polarisation waveforms}
\label{sec6}

The asymptotic waveform at future null infinity from an isolated source is the transverse-traceless (TT) projection of the metric deviation at the leading order $1/R$ in the distance $R=\vert\mathbf{X}\vert$ to the source, in a radiative coordinate system $X^\mu=(c\,T,\mathbf{X})$.\footnote{Radiative coordinates $T$ and $\mathbf{X}$, also called Bondi-type coordinates \cite{BBM62}, are such that the metric coefficients admit an expansion when $R\rightarrow +\infty$ with $U\equiv T-R/c$ being constant, in simple powers of $1/R$, without the logarithms of $R$ plaguing the harmonic coordinate system. Here $U$ is a null or asymptotically null characteristic. It is known that the ``far-zone'' logarithms in harmonic coordinates can be removed order-by-order by going to radiative coordinates \cite{B87}.} The waveform can be uniquely decomposed \cite{Th80} into radiative multipole components parametrized by mass-type moments $\mathrm{U}_L$ and current-type ones $\mathrm{V}_L$. We shall define the radiative moments in such a way thay they agree with the $\ell$-th time derivatives of the source moments $\mathrm{I}_L$ and $\mathrm{J}_L$ at the linear level, \textit{i.e.}
\begin{subequations}\label{UVIJ}
\begin{align}
\mathrm{U}_L &= \mathrm{I}_L^{(\ell)} + \mathcal{O}(G)\,,\\
\mathrm{V}_L &= \mathrm{J}_L^{(\ell)} + \mathcal{O}(G)\,.
\end{align}
\end{subequations}
At the non-linear level the radiative moments will crucially differ from the source moments; the relations between these two types of moments will be discussed in the next Section. The radiative moments $\mathrm{U}_L(U)$ and $\mathrm{V}_L(U)$ are functions of the retarded time $U\equiv T-R/c$ in radiative coordinates.

The asymptotic waveform at distance $R$ and retarded time $U$ is then given by
\begin{align}\label{hijTT}
h^\mathrm{TT}_{ij} &= \frac{4G}{c^2R} \,\mathcal{P}_{ijkl}
\sum^{+\infty}_{\ell=2}\frac{1}{c^\ell \ell !} \biggl\{ N_{L-2}
\,\mathrm{U}_{klL-2} - \frac{2\ell}{c(\ell+1)} \,N_{aL-2} \,\varepsilon_{ab(k}
\,\mathrm{V}_{l)bL-2} \biggr\} \,.
\end{align}
We denote by $\mathbf{N} = \mathbf{X}/R = (N_i)$ the unit vector pointing from the source to the far-away detector. The TT projection operator reads $\mathcal{P}_{ijkl} = \mathcal{P}_{ik}\mathcal{P}_{jl}-\frac{1}{2}\mathcal{P}_{ij}\mathcal{P}_{kl}$ where $\mathcal{P}_{ij}=\delta_{ij}-N_iN_j$ is the projector orthogonal to the unit direction $\mathbf{N}$. We project out the asymptotic waveform \eqref{hijTT} on polarization directions in a standard way. We denote the two unit polarisation vectors by $\mathbf{P}$ and $\mathbf{Q}$, which are orthogonal and transverse to the direction of propagation $\mathbf{N}$ (hence $\mathcal{P}_{ij}=P_iP_j+Q_iQ_j$). Our conventions and choice for $\mathbf{P}$ and $\mathbf{Q}$ will be specified in Section \ref{sec12}. Then the two ``plus'' and ``cross'' polarisation states of the waveform are
\begin{subequations}\label{hpc}
\begin{align}
h_+ &\equiv \frac{P_iP_j-Q_iQ_j}{2}\,h^\mathrm{TT}_{ij}\,,\\
h_\times &\equiv \frac{P_iQ_j+P_jQ_i}{2} \,h^\mathrm{TT}_{ij}\,.
\end{align}
\end{subequations}

Although the multipole decomposition \eqref{hijTT} entirely describes the waveform, it is also important, especially having in mind the comparison between the post-Newtonian results and numerical relativity \cite{BCP07}, to consider separately the modes $(\ell,m)$ of the waveform as defined with respect to a basis of spin-weighted spherical harmonics. To this end we decompose $h_+$ and $h_\times$ as (see \textit{e.g.} \cite{BCP07, K07})
\begin{equation}\label{spinw}
h_+ - \ui h_\times = \sum^{+\infty}_{\ell=2}\sum^{\ell}_{m=-\ell} h^{\ell
m}\,Y^{\ell m}_{-2}(\Theta,\Phi)\,,
\end{equation}
where the spin-weighted spherical harmonics of weight $-2$ is a function of the spherical angles $(\Theta,\Phi)$ defining the direction of propagation $\mathbf{N}$ and reads
\begin{subequations}\label{harm}
\begin{align}
Y^{\ell m}_{-2} &= \sqrt{\frac{2\ell+1}{4\pi}}\,d^{\,\ell
m}(\Theta)\,e^{\ui \,m \,\Phi}\,,\\d^{\,\ell m} &\equiv
\sum_{k=k_1}^{k_2}D^{\,\ell m}_k\left(\cos\frac{\Theta}{2}\right)^{\!2\ell+m-2k-2}
\!\!\!\left(\sin\frac{\Theta}{2}\right)^{\!2k-m+2}\,,\\ D^{\,\ell m}_k &\equiv \frac{(-)^k}{k!}
\frac{\sqrt{(\ell+m)!(\ell-m)!(\ell+2)!(\ell-2)!}}
{(k-m+2)!(\ell+m-k)!(\ell-k-2)!}\,.
\end{align}
\end{subequations}
Here $k_1=\mathrm{max}(0,m-2)$ and $k_2=\mathrm{min}(\ell+m,\ell-2)$. Using the orthonormality properties of these harmonics we obtain the separate modes $h^{\ell m}$ from the surface integral (with the overline denoting the complex conjugate)
\begin{equation}\label{decomp}
h^{\ell m} = \int \ud\Omega \,\Bigl[h_+ - \ui h_\times\Bigr] \,\overline{Y}^{\,\ell
m}_{-2} (\Theta,\Phi)\,.
\end{equation}
On the other hand, we can also write $h^{\ell m}$ directly in terms of the radiative multipole moments $\mathrm{U}_L$ and $\mathrm{V}_L$, with result
\begin{equation}\label{inv}
h^{\ell m} = -\frac{G}{\sqrt{2}\,R\,c^{\ell+2}}\left[\mathrm{U}^{\ell
m}-\frac{\ui}{c}\mathrm{V}^{\ell m}\right]\,,
\end{equation}
where $\mathrm{U}^{\ell m}$ and $\mathrm{V}^{\ell m}$ are the radiative mass and current moments in non-STF guise. These are given in terms of the STF moments by
\begin{subequations}\label{UV}
\begin{align}
\mathrm{U}^{\ell m} &= \frac{4}{\ell!}\,\sqrt{\frac{(\ell+1)(\ell+2)}{2\ell(\ell-1)}}
\,\alpha_L^{\ell m}\,\mathrm{U}_L\,,\\ \mathrm{V}^{\ell m} &=
-\frac{8}{\ell!}\,\sqrt{\frac{\ell(\ell+2)}{2(\ell+1)(\ell-1)}}
\,\alpha_L^{\ell m}\,\mathrm{V}_L\,.
\end{align}
\end{subequations}
Here $\alpha_L^{\ell m}$ denotes the STF tensor connecting together the usual basis of spherical harmonics $Y^{\ell m}$ to the set of STF tensors $\hat{N}_L\equiv\mathrm{STF}(N_L)$, recalling that $Y^{\ell m}$ and $\hat{N}_L$ represent two basis of an irreducible representation of weight $\ell$ of the rotation group. They are related by
\begin{subequations}\label{NY}
\begin{align}
\hat{N}_L(\Theta,\Phi) &= \sum_{m=-\ell}^{\ell}
\alpha_L^{\ell m}\,Y^{\ell m}(\Theta,\Phi)\,,\\Y^{\ell m}(\Theta,\Phi) &=
\frac{(2\ell+1)!!}{4\pi \ell!}\,\overline{\alpha}_L^{\ell m}\,\hat{N}_L(\Theta,\Phi)\,,
\end{align}
\end{subequations}
with the STF tensorial coefficient being
\begin{align}\label{alpha}
\alpha_L^{\ell m} &= \int \ud\Omega\,\hat{N}_L\,\overline{Y}^{\,\ell
m}\,.
\end{align}
The decomposition in spherical harmonic modes is especially useful if some of the radiative moments are known to higher post-Newtonian order than others. In this case the comparison with the numerical calculation \cite{BCP07, K07} can be made for these individual modes with higher post-Newtonian accuracy.

\section{Radiative moments versus source moments}
\label{sec7}

The basis of our computation is the general solution of the Einstein field equations outside an isolated matter system computed iteratively in the form of a post-Minkowskian or non-linearity expansion \eqref{MhabPM} (see details in \cite{BD86, BD92}). Here we give some results concerning the relation between the set of radiative moments $\{\mathrm{U}_L, \mathrm{V}_L\}$ and the sets of source moments $\{\mathrm{I}_L, \mathrm{J}_L\}$ and gauge moments $\{\mathrm{W}_L, \cdots, \mathrm{Z}_L\}$. Complete results up to 3PN order are available and have recently been used to control the 3PN waveform of compact binaries \cite{BFIS08}.

Armed with definitions for all those moments, we proceed in a modular way. We express the radiative moments $\{\mathrm{U}_L, \mathrm{V}_L\}$ in terms of some convenient intermediate constructs $\{\mathrm{M}_L, \mathrm{S}_L\}$ called the canonical moments. Essentially these canonical moments take into account the effect of the gauge transformation present in \eqref{hk}. Therefore they differ from the source moments $\{\mathrm{I}_L, \mathrm{J}_L\}$ only at non-linear order. We shall see that in terms of a post-Newtonian expansion the canonical and source moments agree with each other up to 2PN order. The canonical moments are then connected to the actual source multipole moments $\{\mathrm{I}_L, \mathrm{J}_L\}$ and $\{\mathrm{W}_L, \cdots, \mathrm{Z}_L\}$. The point of the above strategy is that the source moments (including gauge moments) admit closed-form expressions as integrals over the stress-energy distribution of matter and gravitational fields in the source, as shown in \eqref{IJL}.

The mass quadrupole moment $\mathrm{U}_{ij}$ (having $\ell=2$) is known up to the 3PN order \cite{B98tail}. At that order it is made of quadratic and cubic non-linearities, and we have
\begin{align}\label{U2}
\mathrm{U}_{ij}(U) &= \mathrm{M}^{(2)}_{ij} (U) + {2G \,\mathrm{M}\over c^3} \int_{-\infty}^{U} \ud
u \left[ \ln \left({U-u\over 2u_0}\right)+{11\over12} \right]
\mathrm{M}^{(4)}_{ij} (u) \nonumber \\ &+{G\over
c^5}\left\{-{2\over7}\int_{-\infty}^{U} \ud u \,\mathrm{M}^{(3)}_{a\langle
i}(u)\mathrm{M}^{(3)}_{j\rangle a}(u) \right.\nonumber \\ &\qquad~\left. + {1
\over7}\mathrm{M}^{(5)}_{a\langle i}\mathrm{M}_{j\rangle a} - {5 \over7} \mathrm{M}^{(4)}_{a\langle
i}\mathrm{M}^{(1)}_{j\rangle a} -{2 \over7} \mathrm{M}^{(3)}_{a\langle i}\mathrm{M}^{(2)}_{j\rangle a}
+{1 \over3}\varepsilon_{ab\langle i}\mathrm{M}^{(4)}_{j\rangle a}\mathrm{S}_{b}\right\}\nonumber
\\ &+ 2\left({G \,\mathrm{M}\over c^3}\right)^2\int_{-\infty}^{U} \ud u \left[ \ln^2
\left({U-u\over 2u_0}\right)+{57\over70} \ln\left({U-u\over
2u_0}\right)+{124627\over44100} \right] \mathrm{M}^{(5)}_{ij} (u) \nonumber \\ &
+\,\, \mathcal{O}\left(\frac{1}{c^7}\right)\,.
\end{align}
Notice the quadratic tail integral at 1.5PN order, the cubic tail-of-tail integral at 3PN order, and the non-linear memory integral at 2.5PN order \cite{Chr91, Th92, WW91, BD92}. The tail is composed of the coupling between the mass quadrupole moment $\mathrm{M}_{ij}$ and the mass monopole moment or total mass $\mathrm{M}$; the tail-of-tail is a coupling between $\mathrm{M}_{ij}$ and two monopoles $\mathrm{M}\times\mathrm{M}$; the non-linear memory is a coupling $\mathrm{M}_{ij}\times\mathrm{M}_{kl}$. All these ``hereditary'' integrals imply a dependence of the waveform on the complete history of the source, from infinite past up to the current retarded time $U\equiv T-R/c$. The constant $u_0$ in the tail integrals is defined by $u_0\equiv r_0/c$, where $r_0$ is the arbitrary length scale introduced in \eqref{rB}.

Note that the dominant hereditary integral is the tail arising at 1.5PN order in all radiative moments. For general $\ell$ we have at that order
\begin{subequations}\label{ULVL}
\begin{align}
    \mathrm{U}_L &= \mathrm{M}^{(\ell)}_L  +
    \frac{2G\,\mathrm{M}}{c^3} \int_{-\infty}^U \ud u \,
    \left[ \ln \left( \frac{U-u}{2u_0}
    \right) + \kappa_\ell \right] \mathrm{M}^{(\ell+2)}_L (u) + \mathcal{O}\left( \frac{1}{c^5} \right)\,,
    \\ \mathrm{V}_L &= \mathrm{S}^{(\ell)}_L 
    + \frac{2G\,\mathrm{M}}{c^3} \int_{-\infty}^U \ud u\,
    \left[ \ln \left( \frac{U-u}{2u_0}
    \right) + \pi_\ell \right] \mathrm{S}^{(\ell+2)}_L (u) + \mathcal{O}\left( \frac{1}{c^5} \right)\,,
\end{align}
\end{subequations}
where the constants $\kappa_\ell$ and $\pi_\ell$ are given by
\begin{subequations}\label{kappapi}
\begin{align}
    \kappa_\ell &= \frac{2\ell^2+5\ell+4}{\ell(\ell+1)(\ell+2)}+\sum_{k=1}^{\ell-2}\frac{1}{k}\,,
    \\ \pi_\ell &= \frac{\ell-1}{\ell(\ell+1)}+\sum_{k=1}^{\ell-1}\frac{1}{k}\,.
  \end{align}
\end{subequations}
Now it can be proved that the retarded time $U$ in radiative coordinates reads
\begin{equation}\label{Utr}
  U=t-\frac{r}{c}-\frac{2G\,\mathrm{M}}{c^3}\ln\left(\frac{r}{r_0}\right) + \mathcal{O}\left(\frac{1}{c^5}\right)\,,
\end{equation}
where $(t, r)$ are the harmonic coordinates. Inserting $U$ into \eqref{ULVL} we obtain the radiative moments expressed in terms of local source-rooted coordinates $(t, r)$, \textit{e.g.}
\begin{equation}
  \mathrm{U}_L = \mathrm{M}^{(\ell)}_L(t-r/c) +
  \frac{2G\,\mathrm{M}}{c^3} \int_{-\infty}^{t-r/c} \!\!\!\!\! \ud u 
  \left[ \ln \left( \frac{t-u-r/c}{2r/c}
    \right) + \kappa_\ell \right] \mathrm{M}^{(\ell+2)}_L (u) + \mathcal{O}\left( \frac{1}{c^5} \right)\,.
\end{equation}
This no longer depends on the constant $u_0$ --- \textit{i.e.} the $u_0$ gets replaced by the retardation time $r/c$. More generally it can be checked that $u_0$ always disappears from physical results at the end. On the other hand we can be convinced that the constant $\kappa_\ell$ (and $\pi_\ell$ as well) depends on the choice of source-rooted coordinates $(t, r)$. For instance if we change the harmonic coordinate system $(t, r)$ to some ``Schwarzschild-like'' coordinates $(t', r')$ such that $t'=t$ and $r'=r+G\mathrm{M}/c^2$, we get a new constant $\kappa'_\ell=\kappa_\ell+1/2$. Thus we have $\kappa_2=11/12$ in harmonic coordinates [as shown in \eqref{U2}], but $\kappa'_2=17/12$ in Schwarzschild coordinates.

We still have to relate the canonical moments $\{\mathrm{M}_L, \mathrm{S}_L\}$ to the source multipole moments. As we said the difference between these two types of moments comes from the gauge transformation \eqref{phia} and arises only at the small 2.5PN order. The consequence is that we have to worry about this difference only for high post-Newtonian waveforms. For the mass quadrupole moment $\mathrm{M}_{ij}$, the requisite correction is given by
\begin{equation}\label{M2}
\mathrm{M}_{ij}= \mathrm{I}_{ij}+\frac{4G}{c^5} \left[\mathrm{W}^{(2)}\mathrm{I}_{ij}-\mathrm{W}^{(1)}\mathrm{I}_{ij}^{(1)}\right] +
\mathcal{O}\left(\frac{1}{c^7}\right)\,,
\end{equation}
where $\mathrm{I}_{ij}$ denotes the source mass quadrupole, and where $\mathrm{W}$ is the monopole corresponding to the gauge moments $\mathrm{W}_L$ (\textit{i.e.} the moment having $\ell=0$). Up to 3PN order, $\mathrm{W}$ is only needed at Newtonian order. The expression \eqref{M2} is valid in a mass-centred frame defined by the vanishing of the conserved mass dipole moment: $\mathrm{I}_i=0$. 

Note that closed-form formulas generalizing \eqref{M2} and similar expressions to all post-Newtonian orders (and all multipole interactions), if they exist are not known; they need to be investigated anew for specific cases. Thus it is convenient in the present approach to systematically keep the source moments $\{\mathrm{I}_L,\mathrm{J}_L,\mathrm{W}_L,\mathrm{X}_L,\mathrm{Y}_L,\mathrm{Z}_L\}$ as the fundamental variables describing the source.

\bigskip\bigskip
\centerline{\large{\bf{B~ Inspiralling compact binaries}}}
\vspace{-0.4cm}
\section{Stress-energy tensor of spinning particles}
\label{sec8}

So far the post-Newtonian formalism has been developed for arbitrary matter distributions. We want now to apply it to material systems made of compact objects (neutron stars or black holes) which can be described with great precision by point masses. We thus discuss the modelling of point-particles possibly carrying some intrinsic rotation or spin. This means finding the appropriate stress-energy tensor which will have to be inserted into the general post-Newtonian formulas such as the expressions of the source moments \eqref{IJL}.

In the general case the stress-energy tensor will be the sum of a ``monopolar'' piece, which is a linear combination of monopole sources, \textit{i.e.} made of Dirac delta-functions, plus the ``dipolar'' or spin piece, made of \textit{gradients} of Dirac delta-functions. Hence we write
\begin{equation}\label{Tab}
T\ab = T\ab_\text{mono} + T\ab_\text{spin}\,.
\end{equation}
The monopole part takes the form of the stress-energy tensor for $N$ particles (labelled by $A=1,\cdots,N$) without spin, reading in a four-dimensional picture
\begin{equation}\label{TabM}
T\ab_\text{mono} = c \sum_{A=1}^N
\int_{-\infty}^{+\infty}\ud\tau_A\,p_A^{(\alpha}\,u_A^{\beta)}\,\frac{\delta^{(4)}
(x-y_A)}{\sqrt{-(g)_A}}\,.
\end{equation}
Here $\delta^{(4)}$ is the four-dimensional Dirac function. The world-line of particle $A$, denoted $y_A^\alpha$, is parametrized by the particle's proper time $\tau_A$. The four-velocity is given by $u_A^\alpha=\ud y_A^\alpha/\ud\tau_A$ and is normalized to $(g_{\mu\nu})_A u_A^\mu u_A^\nu = -c^2$, where $(g_{\mu\nu})_A$ denotes the metric at the particle's location. The four-vector $p_A^\alpha$ is the particle's linear momentum. For particles without spin we shall simply have $p_A^\alpha = m_A u_A^\alpha$. However for spinning particles $p_A^\alpha$ will differ from that and include some contributions from the spins as given by \eqref{pu} below.

The dipolar part of the stress-energy tensor depends specifically on the spins, and reads in the classic formalism of spinning particles (due to Tulczyjew \cite{Tulc1, Tulc2}, Trautman \cite{Traut58}, Dixon \cite{Dixon}, Bailey \& Israel \cite{BI80}),
\begin{equation}\label{TabS}
T\ab_\text{spin} =-\sum_{A=1}^N\nabla_\mu\biggl[\int_{-\infty}^{
+\infty}\ud\tau_A\, S_A^{\mu(\alpha}\,u_A^{\beta)}\,\frac{\delta^{(4)}
(x-y_A)}{\sqrt{-(g)_A}}\biggr]\,,
\end{equation}
where $\nabla_\mu$ is the covariant derivative, and the anti-symmetric tensor $S_A\ab$ represents the spin angular momentum of particle $A$. In this formalism the momentum-like quantity $p_A^\alpha$ [entering \eqref{TabM}] is a time-like solution of the equation
\begin{equation}\label{momentum}
\frac{\uD S_A\ab}{\ud\tau_A} = \left(p_A^\alpha u_A^\beta - p_A^\beta u_A^\alpha\right)\,,
\end{equation}
where $\uD/\ud\tau_A$ denotes the covariant proper time derivative. The equation of translational motion of the spinning particle, equivalent to the covariant conservation $\nabla_\mu T^{\alpha\mu}=0$ of the total stress-energy tensor \eqref{Tab}, is the Papapetrou equation \cite{Papa51spin} involving a coupling to curvature,
\begin{equation} \label{eom}
\frac{\uD p_A^\alpha}{\ud\tau_A} = -\frac{1}{2} \,S_A^{\mu\nu} \,u_A^\rho \,(R^\alpha_{~\rho\mu\nu})_A \,.
\end{equation}
The Riemann tensor is evaluated at the particle's position $A$. This equation can also be derived directly from an action principle \cite{BI80}.

It is well-known that a choice must be made for a supplementary spin condition in order to fix unphysical degrees of freedom associated with an arbitrariness in the definition of the spin tensor $S\ab$. This arbitrariness can be interpreted, in the case of extended bodies, as a freedom in the choice for the location of the center-of-mass worldline of the body, with respect to which the angular momentum is defined (see \textit{e.g.} \cite{K95}). An elegant spin condition is the covariant one,
\begin{equation}\label{SSC}
S_A^{\alpha\mu}\,p^A_\mu = 0 \,,
\end{equation}
which allows a natural definition of a spin four-(co)vector $S^A_\alpha$ such that
\begin{equation}\label{spin4}
S_A\ab = -\frac{1}{\sqrt{-(g)_A}}\,\varepsilon^{\alpha\beta\mu\nu}\,
\frac{p^A_\mu}{m_A c} \,S^A_\nu \,.
\end{equation}
For the spin vector $S^A_\alpha$ itself, we can choose a four-vector which is purely spatial in the particle's instantaneous rest frame, where $u_A^\alpha=(1,\mathbf{0})$. Therefore, we deduce that in any frame
\begin{equation}\label{Su}
S^A_\mu u_A^\mu = 0 \,.
\end{equation}
As a consequence of the covariant spin condition \eqref{SSC}, we easily verify that the spin scalar is conserved along the trajectories, \textit{i.e.}
\begin{equation}\label{spinscalar}
S_A^{\mu\nu} S^A_{\mu\nu} = \mathrm{const}\,.
\end{equation}
Furthermore, we can check, using \eqref{SSC} and also the law of motion \eqref{eom}, that the mass defined by $m_A^2 c^2 = -p_A^\mu p^A_\mu$ is also constant along the trajectories: $m_A=\mathrm{const}$. Finally, the relation linking the four-momentum $p_A^\alpha$ and the four-velocity $u_A^\alpha$ is readily deduced from the contraction of \eqref{momentum} with the four-momentum, which results in
\begin{equation}\label{pu}
p_A^\alpha (pu)_A + m_A^2\,c^2\,u_A^\alpha = \frac{1}{2} \,S_A^{\alpha\mu}
\,S_A^{\nu\rho} \,u_A^\sigma \,(R_{\mu\sigma\nu\rho})_A\,,
\end{equation}
where $(pu)_A\equiv p^A_\mu u_A^\mu$. Contracting further this relation with the four-velocity one deduces the expression of $(pu)_A$ and inserting this back into~\eqref{pu} yields the desired relation between $p_A^\alpha$ and $u_A^\alpha$.  

Focusing our attention on spin-orbit interactions, which are \textit{linear} in the spins, we can neglect quadratic and higher spin corrections denoted $\mathcal{O}(S^2)$; drastic simplifications of the formalism occur in this case. Since the right-hand-side of \eqref{pu} is quadratic in the spins, we find that the four-momentum is linked to the four-velocity by the simple proportionality relation
\begin{equation}\label{pulin}
p_A^\alpha = m_A\,u_A^\alpha + \mathcal{O}(S^2)\,.
\end{equation}
Hence, the spin condition \eqref{SSC} becomes
\begin{equation}\label{SSClin}
S_A^{\alpha\mu}\,u^A_\mu = \mathcal{O}(S^3)\,.
\end{equation}
Also, the equation of evolution for the spin, sometimes called the \textit{precessional} equation, follows immediately from the relationship \eqref{momentum} together with the law \eqref{pulin} as 
\begin{equation}\label{parallel}
\frac{\uD S_A\ab}{\ud\tau_A} = \mathcal{O}(S^2)\quad\Longleftrightarrow\quad\frac{\uD S^A_\alpha}{\ud \tau_A} = \mathcal{O}(S^2)\,.
\end{equation}
Hence the spin vector $S^A_\alpha$ satisfies the equation of parallel transport, which means that it remains constant in a freely falling frame, as could have been expected beforehand. Of course the norm of the spin vector is preserved along trajectories, 
\begin{equation}\label{spinnorm}
S^A_\mu S_A^\mu = \mathrm{const}\,.
\end{equation}
In Section \ref{sec13} we shall apply this formalism to the study of spin-orbit effects in the equations of motion and energy flux of compact binaries.

\section{Hadamard regularization}
\label{sec9}

The stress-energy tensor of point masses has been defined in the previous Section by means of Dirac functions, and involves metric coefficients evaluated at the locations of the point particles, namely $(g\lab)_A$. However, it is clear that the metric $g\lab$ becomes \textit{singular} at the particles. Indeed this is already true at Newtonian order where the potential generated by $N$ particles reads
\begin{equation}\label{U}
U = \sum_{B=1}^N \frac{G m_B}{r_B}\,,
\end{equation}
where $r_B\equiv\vert\mathbf{x}-\mathbf{y}_B\vert$ is the distance between the field point and the particle $B$. Thus the values of $U$ and hence of the metric coefficients [recall that $g_{00} = -1 +2U/c^2+\mathcal{O}(c^{-4})$], are ill-defined at the locations of the particles. What we need is a \textit{self-field regularization}, \textit{i.e.} a prescription for removing the infinite self-field of the point masses. Arguably the choice of a particular regularization constitutes a fully qualified element of our physical modelling of compact objects. At Newtonian order the regularization of the potential \eqref{U} should give the well-known result
\begin{equation}\label{UA}
(U)_A = \sum_{B\not=A} \frac{G m_B}{r_{AB}}\,,
\end{equation}
where $r_{AB}\equiv\vert\mathbf{y}_A-\mathbf{y}_B\vert$ and the infinite self interaction term has simply been discarded from the summation. At high post-Newtonian orders the problem is not trivial and the self-field regularization must be properly defined.

The post-Newtonian formalism reviewed in Sections \ref{sec2}--\ref{sec7} assumed from the start a continuous (smooth) matter distribution. Actually this formalism will be applicable to singular point-mass sources, described by the stress-energy tensor of Section \ref{sec8}, provided that we suplement the scheme by the self-field regularization. Note that this regularization has nothing to do with the finite-part process $\FP$ extensively used in the case of extended matter sources. The latter finite part was an ingredient of the rigorous derivation of the general post-Newtonian solution [see \eqref{genpoiss}], while the self-field regularization is an assumption regarding a particular type of singular source.

Our aim is to compute up to 3PN order the metric coefficients at the location of one of the particles: $(g_{\alpha\beta})_A$. At this stage different self-field regularizations are possible. We first review Hadamard's regularization \cite{Hadamard, Schwartz}, which has proved to be very efficient for doing practical computations, but suffers from the important drawback of yielding some ``ambiguity parameters'', which cannot be determined within the regularization, starting at the 3PN order.

Iterating the Einstein field equations with point-like matter sources (delta functions with spatial supports localized on $\mathbf{y}_A$) yields a generic form of functions representing the metric coefficients in successive post-Newtonian approximations. The generic functions, say $F(\mathbf{x})$, are smooth except at the points $\mathbf{y}_A$, around which they admit singular Laurent expansions in powers and inverse powers of $r_A\equiv\vert\mathbf{x}-\mathbf{y}_A\vert$. When $r_A\rightarrow 0$ we have (say, for any $P\in\N$)
\begin{equation}\label{Fx}
F(\mathbf{x})=\sum_{p=p_0}^P r_A^{p}
\mathop{f}_A{}_{\!\!p}(\mathbf{n}_A)+o(r_A^P)\,.
\end{equation}
The coefficients ${}_Af_p$ of the various powers of $r_A$ depend on the unit direction $\mathbf{n}_A\equiv(\mathbf{x}-\mathbf{y}_A)/r_A$ of approach to the singular point $A$. The powers $p$ are relative integers, and are bounded from below by $p_0\in\Z$. The Landau $o$-symbol for remainders takes its standard meaning. The ${}_Af_p$'s depend also on the (coordinate) time $t$, through their dependence on velocities $\mathbf{v}_B(t)$ and relative positions $\mathbf{y}_{BC}(t)\equiv\mathbf{y}_{B}(t)-\mathbf{y}_{C}(t)$; however the time $t$ is purely ``spectator'' in the regularization process, and thus will not be indicated. The coefficients ${}_Af_p$ for which $p<0$ are referred to as the \textit{singular} coefficients of $F$ around $A$.

The function $F$ being given that way, we define the \textit{Hadamard partie finie} as the following value of $F$ at the location of the particle $A$,\footnote{With this definition it is immediate to check that the previous Newtonian result \eqref{UA} will hold.} 
\begin{equation}\label{FA}
  (F)_A= \langle\mathop{f}_{A}{}_{\!\!0}\rangle\equiv\int \frac{\ud\Omega_A}{ 4\pi}\,\mathop{f}_{A}{}_{\!\!0}(\mathbf{n}_A)\,,
\end{equation}
where $\ud\Omega_A$ denotes the solid angle element centred on $\mathbf{y}_A$ and sustained by $\mathbf{n}_A$. The brackets $\langle\rangle$ mean the angular average. The second notion of Hadamard partie finie concerns the integral $\int \ud^3\mathbf{x}\,F$, which is generically divergent at the points $\mathbf{y}_A$. Its partie finie (in short $\mathrm{Pf}$) is defined by
\begin{equation}\label{PfF}
  \mathrm{Pf}_{s_1\cdots s_N} \int \ud^3\mathbf{x} \, F = \lim_{s \rightarrow 0}
  \, \biggl\{\int_{\mathbb{R}^3\setminus\bigcup B_A(s)} \ud^3\mathbf{x} \, F + 4\pi\sum_{A=1}^N D_A(s)  \biggr\}\,.
\end{equation}
The first term integrates over the domain $\mathbb{R}^3\setminus\bigcup_{A=1}^N B_A(s)$ defined as $\mathbb{R}^3$ deprived from the $N$ spherical balls $B_A(s)\equiv\{\mathbf{x};\,r_A\leq s\}$ of radius $s$ and centred on the points $\mathbf{y}_A$. The second term is the opposite of the sum of divergent parts associated with the first term around each of the particles in the limit where $s\rightarrow 0$. We have
\begin{equation}\label{DAs}
  D_A(s) = \sum_{p=p_0}^{-4}\,\frac{s^{p+3}}{p+3}\,\langle\mathop{f}_{A}{}_{\!\!p}\rangle + \ln
  \left(\frac{s}{s_A}\right) \langle\mathop{f}_{A}{}_{\!\!-3}\rangle\,.
\end{equation}
Since the divergent parts are cancelled (by definition) the Hadamard partie finie is obtained in the limit $s\rightarrow 0$. Notice that as indicated in \eqref{PfF} the Hadamard partie-finie integral is not fully specified: it depends on $N$ strictly positive and \textit{a priori} arbitrary constants $s_1, \cdots, s_N$ parametrizing the logarithms in \eqref{DAs}. 

We have seen that the post-Newtonian scheme consists of breaking the hyperbolic d'Alembertian operator $\Box$ into the elliptic Laplacian $\Delta$ and the retardation term $c^{-2}\partial_t^2$ considered to be small, and put in the right-hand-side of the equation where it can be iterated; see \eqref{Poisson}. We thus have to deal with the regularization of Poisson integrals, or iterated Poisson integrals, of some generic function $F$. The Poisson integral will be divergent\footnote{We consider only the local divergencies due to the singular points $\mathbf{y}_A$. The problem of divergencies of Poisson integrals at infinity is part of the general post-Newtonian formalism and has been treated in Section \ref{sec3}.} and we apply the prescription \eqref{PfF}. Thus,
\begin{equation}\label{Px}
P(\mathbf{x}')= -\frac{1}{4\pi}\,\mathrm{Pf}_{s_1\cdots s_N} \int\frac{\ud^3{\mathbf{x}}}{\vert{\mathbf{x}}-\mathbf{x}'\vert}
F({\mathbf{x}})\,.
\end{equation}
This definition is valid for each field point $\mathbf{x}'$ \textit{different} from the $\mathbf{y}_A$'s, and we want to investigate the singular limit when $\mathbf{x}'$ tends to one of the source points $\mathbf{y}_A$, so as to define the object $(P)_A$. The definition \eqref{FA} is not directly applicable because the expansion of the Poisson integral $P(\mathbf{x}')$ when $\mathbf{x}'\rightarrow\mathbf{y}_A$ will involve besides the normal powers of $r_A'\equiv\vert\mathbf{x}'-\mathbf{y}_A\vert$ some \textit{logarithms} of $r_A'$. The proper way to define the Hadamard partie finie in this case is to include the $\ln r'_A$ into the definition \eqref{FA} as if it were a mere constant parameter. With this definition we arrive at \cite{BFreg}
\begin{equation}\label{PA}
(P)_A = -\frac{1}{4\pi}\,\mathrm{Pf}_{s_1\cdots s_N}
\int\frac{\ud^3{\mathbf{x}}}{r_A} F({\mathbf{x}}) + \left[\ln\left(\frac{r'_A}{s_A}\right)-1\right]\langle\mathop{f}_{A}{}_{\!\!-2}\rangle\,.
\end{equation}
The first term is given by a partie-finie integral following the definition \eqref{PfF}; the second involves the logarithm of $r'_A$. The constants $s_1, \cdots, s_N$ come from \eqref{Px}. Since $r'_A$ is actually tending to zero, $\ln r'_A$ represents a formally infinite ``constant'', which will ultimately parametrize the final Hadamard regularized 3PN equations of motion. In the two-body case we shall find that the constants $r'_A$ are unphysical in the sense that they can be removed by a coordinate transformation \cite{BFeom}. Note that the apparent dependence of \eqref{PA} on the constant $s_A$ is illusory. Indeed the dependence on $s_A$ cancels out between the first and the second terms in the right-hand-side of \eqref{PA}, so the result depends only on $r_A'$ and the $s_B$'s for $B\not=A$. We thus have a simpler rewriting of \eqref{PA} as
\begin{equation}\label{PA'}
(P)_A = -\frac{1}{4\pi}\,\mathrm{Pf}_{s_1\cdots r'_A\cdots s_N}
\int\frac{\ud^3{\mathbf{x}}}{r_A} F({\mathbf{x}}) - \langle\mathop{f}_{A}{}_{\!\!-2}\rangle\,.
\end{equation}
Unfortunately, the constants $s_B$ for $B\not=A$ remaining in the result \eqref{PA'} will be the source of a genuine ambiguity. This ambiguity can in fact be traced back to the so-called ``non-distributivity'' of the Hadamard partie finie, a consequence of the presence of the angular integration in \eqref{FA}, and implying that $(FG)_A\not= (F)_A(G)_A$ in general. The non-distributivity arises precisely at the 3PN order both in the equations of motion and radiation field of point-mass binaries. At that order we are loosing with Hadamard's regularization an elementary rule of ordinary calculus. Consequently we expect that some basic symmetries of general relativity such as diffeomorphism invariance will be lost. However Hadamard's regularization can still be efficiently used to compute most of the terms in the equations of motion and radiation field at 3PN order; only a few ambiguous terms will show up which have then to be determined by another method.

\section{Dimensional regularization}
\label{sec10}

Dimensional regularization is an extremely powerful regularization which is free of ambiguities (at least up to the 3PN order). The main reason is that it is able to preserve the symmetries of classical general relativity; in fact dimensional regularization was invented \cite{tHooft, Bollini} as a means to preserve the gauge symmetry of perturbative quantum field theories. In the present context we shall show that dimensional regularization permits to resolve the problem of ambiguities arising at the 3PN order in Hadamard's regularization. We shall employ dimensional regularization not merely as a trick to compute some particular integrals which would otherwise be divergent or ambiguous, but as a fundamental tool for solving in a consistent way the Einstein field equations with singular sources. We therefore assume that the correct theory is general relativity in $D=d+1$ space-time dimensions. As usual, any intermediate formulas will be interpreted by analytic continuation for a general complex spatial dimension $d\in\C$. In particular we shall analytically continue $d$ down to the value of interest 3 and pose
\begin{equation}
d = 3 + \varepsilon\,.
\end{equation}
The Einstein field equations in $d$ spatial dimensions take the same form as presented in Section \ref{sec2}, with the exception that the explicit expression of the gravitational source term $\Lambda^{\alpha\beta}$ now depends on $d$. We find that only the last term in \eqref{Lambda} acquires a dependence on $d$; namely the factor $\frac{1}{2}$ in $g_{\rho\sigma}g_{\epsilon\pi}-\frac{1}{2} g_{\sigma\epsilon}g_{\rho\pi}$ should now read $\frac{1}{d-1}$. In addition the $d$-dimensional gravitational constant is related to the usual three-dimensional Newton constant $G$ by
\begin{equation}\label{Geps}
G^{(d)} = G\,\ell_0^\varepsilon\,,
\end{equation}
where $\ell_0$ is a characteristic length associated with dimensional regularization. 

In the post-Newtonian iteration performed in $d$ dimensions we shall meet the analogue of the function $F$, which we denote by $F^{(d)}(\mathbf{x})$ where $\mathbf{x}\in\R^d$. It turns out that in the vicinity of the singular points $\mathbf{y}_A$, the function $F^{(d)}$ admits an expansion richer than in \eqref{Fx}, and of the type
\begin{equation}\label{Fdx}
F^{(d)}(\mathbf{x})=\sum_{p=p_0}^P \sum_{q=q_0}^{q_1}r_A^{p+q\varepsilon}\mathop{f}_A{}_{\!\!p,q}^{(\varepsilon)}(\mathbf{n}_A)+o(r_A^P)\,.
\end{equation}
The coefficients ${}_Af_{p,q}^{(\varepsilon)}(\mathbf{n}_1)$ depend on the dimension through $\varepsilon\equiv d-3$ and also on the scale $\ell_0$. The powers of $r_A$ are now of the type $p+q\varepsilon$ where the two relative integers $p,\,q\in\Z$ have values limited as indicated. Because $F^{(d)}$ reduces to $F$ when $\varepsilon=0$ we necessarily have the constraints
\begin{equation}\label{fdf}
\sum_{q=q_0}^{q_1}\,\mathop{f}_A{}_{\!\!p,q}^{(0)} = \mathop{f}_A{}_{\!\!p}\,.
\end{equation}

To proceed with the iteration we need the Green function of the Laplace operator in $d$ dimensions. Its explicit form is
\begin{equation}\label{uA}
u_A^{(d)} = K\,r_A^{2-d}\,.
\end{equation}
It satisfies $\Delta u_A^{(d)} = -4\pi\,\delta_A^{(d)}$, where $\delta_A^{(d)} \equiv \delta^{(d)}(\mathbf{x}-\mathbf{y}_A)$ denotes the $d$-dimensional Dirac function. The constant $K$ is given by
\begin{equation}\label{K}
K=\frac{\Gamma\left(\frac{d-2}{2}\right)}{\pi^{
\frac{d-2}{2}}}\,,
\end{equation}
where $\Gamma$ is the usual Eulerian function. It reduces to one when $d\rightarrow 3$. Note that the volume $\Omega_{d-1}$ of the sphere with $d-1$ dimensions is related to $K$ by
\begin{equation}\label{sphere}
\Omega_{d-1}=\frac{4\pi}{(d-2)K}\,.
\end{equation}
With these results the Poisson integral of $F^{(d)}$, constituting the $d$-dimensional analogue of \eqref{Px}, reads
\begin{equation}\label{Pdx}
P^{(d)}(\mathbf{x}')= -\frac{K}{4\pi}
\int\frac{\ud^d{\mathbf{x}}}{\vert{\mathbf{x}}-\mathbf{x}'\vert^{d-2}}\,F^{(d)}({\mathbf{x}})\,.
\end{equation}
In dimensional regularization the singular behavior of this integral is automatically taken care of by analytic continuation in $d$. Next we evaluate the integral at the singular point $\mathbf{x}' = \mathbf{y}_A$. In contrast with Hadamard's regularization where the result was given by \eqref{PA}, in dimensional regularization this is quite easy, as we are allowed to simply replace $\mathbf{x}'$ by $\mathbf{y}_A$ into the explicit integral form \eqref{Pdx}. So we simply have
\begin{equation}\label{PdA}
P^{(d)}({\mathbf{y}}_A)=-\frac{K}{4\pi}
\int\frac{\ud^d{\mathbf{x}}}{r_A^{d-2}}\,F^{(d)}({\mathbf{x}})\,.
\end{equation}

The main step of our strategy \cite{DJSdim, BDE04} will now consist of computing the \textit{difference} between the $d$-dimensional Poisson potential \eqref{PdA} and its $3$-dimensional counterpart which is defined from Hadamard's regularization as \eqref{PA}. We shall then add this difference (in the limit $\varepsilon=d-3 \rightarrow 0$) to the result obtained by Hadamard regularization in order to get the corresponding dimensional regularization result. This strategy is motivated by the fact that as already mentionned most of the terms do not present any problems and have already been correctly computed using Hadamard's regularization. Denoting the difference between the two regularizations by means of the script letter $\mathcal{D}$, we write
\begin{equation}\label{DPA}
\mathcal{D}(P)_A\equiv P^{(d)}({\mathbf{y}}_A)-(P)_A\,.
\end{equation}
We shall only compute the first two terms of the Laurent expansion of $\mathcal{D}(P)_A$ when $\varepsilon\rightarrow 0$, which will be of the form $\mathcal{D}(P)_A = a_{-1}\,\varepsilon^{-1} + a_0 + \mathcal{O}(\varepsilon)$. This is the information needed to determine the value of the ambiguity parameters. Notice that the difference $\mathcal{D}(P)_A$ comes exclusively from the contribution of terms developing some \textit{poles} $\propto 1/\varepsilon$ in the $d$-dimensional calculation. The ambiguity in Hadamard's regularization at 3PN order is reflected by the appearance of poles in $d$ dimensions. The point is that in order to obtain the difference $\mathcal{D}(P)_A$ we do not need the expression of $F^{(d)}$ for an arbitrary source point $\mathbf{x}$ but only in the vicinity of the singular points $\mathbf{y}_A$. Thus this difference depends only on the singular coefficients of the local expansions of $F^{(d)}$ near the singularities. We find \cite{BDE04}
\begin{align}\label{DPAexpl}
\mathcal{D}(P)_A=& -\frac{1}{\varepsilon (1+\varepsilon)}\sum_{q=q_0}^{q_1}\Biggl[\left(\frac{1}{q}+\varepsilon \Bigl[\ln
r_A'-1\Bigr]\right)\langle\mathop
{f}_A{}_{-2,q}^{(\varepsilon)}\rangle\nonumber\\
& + \sum_{B\not=A}\left(\frac{1}{q+1}+\varepsilon\ln s_B\right)
\sum_{\ell=0}^{+\infty}\frac{(-)^\ell}{\ell!}\partial_L
\biggl(\frac{1}{r_{AB}^{1+\varepsilon}}\biggr)\langle n_B^L\,\mathop{f}_B{}_{-\ell-3,q}^{(\varepsilon)}\rangle\Biggr] +\mathcal{O}(\varepsilon)\,.
\end{align}
We still use the bracket notation to denote the angular average but this time performed in $d$ dimensions, \textit{i.e.}
\begin{equation}\label{anguld}
\langle\mathop{f}_A{}_{p,q}^{(\varepsilon)}\rangle
\equiv\int\frac{\ud\Omega_{d-1}(\mathbf{n}_A)}{\Omega_{d-1}}
\mathop{f}_A{}_{p,q}^{(\varepsilon)}(\mathbf{n}_A)\,.
\end{equation}

The above differences for all the Poisson and interated Poisson integrals composing the equations of motion (\textit{i.e.} the accelerations of the point masses) are added to the corresponding results of the Hadamard regularization in the variant of it called the ``pure Hadamard-Schwartz'' regularization (see \cite{BDE04} for more details). In this way we find that the equations of motion in dimensional regularization are composed of a pole part $\propto 1/\varepsilon$ which is purely 3PN, followed by a finite part when $\varepsilon\rightarrow 0$, plus the neglected terms $\mathcal{O}(\varepsilon)$. It has been shown (in the two-body case $N=2$) that:
\begin{enumerate}
\item The pole part $\propto 1/\varepsilon$ of the accelerations can be renormalized into some \textit{shifts} of the ``bare'' world-lines by $\mathbf{y}_A\rightarrow \mathbf{y}_A+\bm{\xi}_A$, with $\bm{\xi}_A$ containing the poles, so that the result expressed in terms of the ``dressed'' world-lines is \textit{finite} when $\varepsilon\rightarrow 0$;
\item The renormalized acceleration is physically equivalent to the result of the Hada-mard regularization, in the sense that it differs from it only by the effects of shifts $\bm{\xi}_A$, if and only if the ambiguities in the Hadamard regularization are fully and uniquely determined (\textit{i.e.} take specific values).
\end{enumerate}
These results \cite{BDE04} provide an unambiguous determination of the equations of motion of compact binaries up to the 3PN order. A related strategy with similar complete results has been applied to the problem of multipole moments and radiation field of point-mass binaries \cite{BDEI04}. This finally completed the derivation of the general relativistic prediction for compact binary inspiral up to 3PN order (and even to 3.5PN order). In later Sections we shall review some features of the 3.5PN gravitational-wave templates of inspiralling compact binaries.

Why should the final results of the employed regularization scheme be unique, in agreement with our expectation that the problem is well-posed and should possess a unique physical answer? The results can be justified by invoking the ``effacing principle'' of general relativity \cite{D83houches} --- namely that the internal structure of the compact bodies does not influence the equations of motion and emitted radiation until a very high post-Newtonian order. Only the masses $m_A$ of the bodies should drive the motion and radiation, and not for instance their ``compactness'' $G m_A/(c^2 a_A)$. A model of point masses should therefore give the correct physical answer, which we expect to be also valid for black holes, provided that the regularization scheme is mathematically consistent.

\section{Energy and flux of compact binaries}
\label{sec11}

The equations of motion of compact binary sources, up to the highest known post-Newtonian order which is 3.5PN, will serve in the definition of the gravitational-wave templates for two purposes:
\begin{enumerate}
\item To compute the center-of-mass energy $E$ appearing in the left-hand-side of the energy balance equation to be used for deducing the orbital phase,
\begin{equation}\label{balance}
\frac{\ud E}{\ud t} = - \mathcal{F}\,;
\end{equation}
\item To \textit{order-reduce} the accelerations coming from the time derivatives of the source multipole moments required to compute the gravitational-wave energy flux $\mathcal{F}$ in the right-hand-side of the balance equation.
\end{enumerate}

We consider two compact objects moving under purely gravitational mutual interaction. In a first stage we assume that the bodies are non-spinning so the motion takes place in a fixed plane, say the x-y plane. The relative position $\mathbf{x} = \mathbf{y}_1 - \mathbf{y}_2$, velocity $\mathbf{v} = \ud\mathbf{x}/\ud t$, and acceleration $\mathbf{a} = \ud\mathbf{v}/\ud t$ are given by
\begin{subequations}\label{xva}\begin{align}
\mathbf{x} &= r \,\mathbf{n} \,,\\ 
\mathbf{v} &= \dot r \,\mathbf{n}
+ r \,\omega \,\bm{\lambda}\,,\\ 
\mathbf{a} &= (\ddot{r} - r
\,\omega^2) \,\mathbf{n} + (r \,\dot{\omega} + 2 \dot{r} \,\omega)\,\bm{\lambda}\,,
\end{align}\end{subequations}
The orbital frequency $\omega$ is related in the usual way to the orbital phase $\phi$ by $\omega = \dot \phi$ (time derivatives are denoted with a dot). Here the vector $\bm{\lambda} = \hat{\mathbf{z}}\times\mathbf{n}$ is perpendicular to the unit vector $\hat{\mathbf{z}}$ along the z-direction orthogonal to the orbital plane, and to the binary's separation unit direction $\mathbf{n}\equiv\mathbf{x}/r$. 

Through 3PN order, it is possible to model the binary's orbit as a \textit{quasi-circular} orbit decaying by the effect of radiation reaction at the 2.5PN order. The restriction to quasi-circular orbits is both to simplify the presentation,\footnote{However the 3PN equations of motion are known in an arbitrary frame and for general orbits.} and for physical reasons because the orbit of inspiralling compact binaries detectable by current detectors should be circular (see the discussion in Section \ref{sec1}). The radiation-reaction effect at 2.5PN order yields\footnote{Mass parameters are the total mass $m\equiv m_1+m_2$, the symmetric mass ratio $\nu\equiv m_1m_2/m^2$ satisfying $0<\nu\leq 1/4$, and for later use the mass difference ratio $\Delta\equiv(m_1-m_2)/m$.\label{footnotemass}}
\begin{subequations}\label{romdot}
\begin{align}
\dot{r} &= - \frac{64}{5} \sqrt{\frac{G m}{r}}~\nu\,\gamma^{5/2} +
\mathcal{O}\left(\frac{1}{c^7}\right)\label{rdot}\,,\\ \dot{\omega} &=
\frac{96}{5} \,\frac{G m}{r^3}\,\nu\,\gamma^{5/2}+
\mathcal{O}\left(\frac{1}{c^7}\right)\label{omdot}\,,
\end{align}
\end{subequations}
where $\gamma$ is defined as the small [\textit{i.e.} $\gamma = \mathcal{O}(c^{-2})$] post-Newtonian parameter
\begin{equation}\label{gamma}
\gamma \equiv {Gm\over rc^2}\,.
\end{equation} 
Substituting these results into \eqref{xva}, we obtain the expressions for the velocity and acceleration during the inspiral,
\begin{subequations}\label{va3PN}\begin{align}
\mathbf{v} &= r \,\omega \,\bm{\lambda} - \frac{64}{5} \sqrt{\frac{G
m}{r}}~\nu\,\gamma^{5/2}\,\mathbf{n} +
\mathcal{O}\left(\frac{1}{c^7}\right)\,,\\ 
\mathbf{a} &= -\omega^2
\,\mathbf{x} - \frac{32}{5}\,\sqrt{\frac{G
m}{r^3}}\,\,\nu\,\gamma^{5/2}\,\mathbf{v} +
\mathcal{O}\left(\frac{1}{c^7}\right)\,.
\end{align}\end{subequations}
Notice that while $\dot{r}=\mathcal{O}(c^{-5})$, we have $\ddot{r}=\mathcal{O}(c^{-10})$ which is of the order of the \textit{square} of radiation-reaction effects and is thus zero with the present approximation.

A central result of post-Newtonian calculations is the expression of the orbital frequency $\omega$ in terms of the binary's separation $r$ up to 3PN order. This result has been obtained independently by three groups. Two are working in harmonic coordinates: Blanchet \& Faye \cite{BFeom, ABF01, BDE04} use a direct post-Newtonian iteration of the equations of motion, while Itoh \& Futamase \cite{IFA01, itoh1, itoh2} apply a variant of the surface-integral approach (\`a la Einstein-Infeld-Hoffmann \cite{EIH}) valid for compact bodies without the need of a self-field regularization. The group of Jaranowski \& Sch\"afer \cite{JaraS98, JaraS99, DJSdim} employs Arnowitt-Deser-Misner coordinates within the Hamiltonian formalism of general relativity.\footnote{This approach is extensively reviewed in the contribution of Gerhard Sch\"afer in this volume.} The 3PN orbital frequency in harmonic coordinates is
\begin{align}\label{omega3PN}
\omega^2 &= {G m\over r^3}\biggl\{ 1+\Bigl(-3+\nu\Bigr)\gamma + \left(6+\frac{41}{4}\nu +\nu^2\right)\gamma^2 \\ &~~+
\left(-10+\left[-\frac{75707}{840}+\frac{41}{64}\pi^2
+22\ln\left(\frac{r}{r'_0}\right) \right]\nu +\frac{19}{2}\nu^2+\nu^3\right)\gamma^3 +
\mathcal{O}\left(\frac{1}{c^8}\right) \biggr\}\nonumber\,.
\end{align}
Note the logarithm at 3PN order coming from a Hadamard self-field regularization scheme, and depending on a constant ${r'}_0$ defined by $m\ln{r'}_0\equiv m_1\ln{r'}_1+m_2\ln{r'}_2$, where $r_A'\equiv\vert\mathbf{x}'-\mathbf{y}_A\vert$ are arbitrary ``constants'' discussed in Section \ref{sec9}. We shall see that ${r'}_0$ disappears from final results --- it can be qualified as a gauge constant.

To obtain the 3PN energy we need to go back to the equations of motion for general non-circular orbits, and deduce the energy as the integral of the motion associated with a Lagrangian formulation of (the conservative part of) these equations \cite{ABF01}. Once we have the energy for general orbits we can reduce it to quasi-circular orbits. We find
\begin{align}\label{constantE}
  E &= -\frac{G m^2\nu}{2 r} \biggl\{ 1 + \left( - \frac{7}{4} +
  \frac{1}{4} \nu \right) \gamma + \left( - \frac{7}{8} + \frac{49}{8}
  \nu + \frac{1}{8} \nu^2 \right) \gamma^2 \\ & \!\!\!+ \left(-\frac{235}{64} 
  + \!\left[\frac{46031}{2240} - \frac{123}{64}
  \pi^2 + \frac{22}{3} \ln \left( \frac{r}{r_0'} \right) \right] \!\nu +
  \frac{27}{32} \nu^2 + \frac{5}{64} \nu^3 \right) \gamma^3  \!+ \mathcal{O} \!\left(\frac{1}{c^8}\right) \biggr\}\,.\nonumber
\end{align}
A convenient post-Newtonian parameter $x = \mathcal{O}(c^{-2})$ is now used in place of $\gamma$; it is defined from the orbital frequency as
\begin{equation}\label{x}
x = \left({G\,m\,\omega \over c^3}\right)^{2/3}\,.
\end{equation}
The interest in this parameter stems from its invariant meaning in a large class of coordinate systems including the harmonic and ADM coordinates. By inverting \eqref{omega3PN} we find at 3PN order 
\begin{align}\label{gammax}
\gamma &= x \biggl\{ 1+\left(1-\frac{\nu}{3}\right)x + 
 \left(1-\frac{65}{12}\nu\right)x^2 \,\\ &~~+ 
 \left(1+\left[-\frac{2203}{2520}-\frac{41}{192}\pi^2
 -\frac{22}{3}\ln\left(\frac{r}{r'_0}\right) \right]\nu
 +\frac{229}{36}\nu^2+\frac{\nu^3}{81}\right)x^3 +
 \mathcal{O}\left(\frac{1}{c^8}\right) \biggr\}\nonumber\,.
\end{align}
This is substituted back into \eqref{constantE} to get the 3PN energy in invariant form. We happily observe that the logarithm and the gauge constant $r'_0$ cancel out in the process and our final result is
\begin{align}\label{Ex}
  E &= -\frac{m \nu c^2 x}{2} \biggl\{ 1 +\left(-\frac{3}{4} -
  \frac{1}{12}\nu\right) x + \left(-\frac{27}{8} +
  \frac{19}{8}\nu -\frac{1}{24}\nu^2\right) x^2
  \nonumber \\
  & + \! \left( \! -\frac{675}{64} +
  \left[\frac{34445}{576} - \frac{205}{96}\pi^2 
  \right]\nu - \frac{155}{96}\nu^2 -
  \frac{35}{5184}\nu^3 \!\right) x^3 + {\cal O}\left(\frac{1}{c^8}\right)\biggr\}\,.
\end{align}

The conserved energy $E$ corresponds to the Newtonian, 1PN, 2PN and 3PN conservative orders in the equations of motion; the damping part is associated with radiation reaction and arises at 2.5PN order. The radiation reaction at the dominant 2.5PN level will correspond to the ``Newtonian'' gravitational-wave flux only. Hence the flying-color 3PN flux $\mathcal{F}$ we are looking for cannot be computed from the 3PN equations of motion alone. Instead we have to apply all the machinery of the post-Newtonian wave generation formalism described in Sections \ref{sec4}--\ref{sec7}. The final result at 3.5PN order is \cite{BIJ02, BFIJ02, BDEI04}
\begin{align}\label{calFx}
\mathcal{F} &= \frac{32c^5}{5G}\nu^2 x^5 \biggl\{ 1 +
  \left(-\frac{1247}{336} - \frac{35}{12}\nu \right) x + 4\pi \,x^{3/2} \nonumber \\
  & \quad \quad \quad +
  \left(-\frac{44711}{9072} + \frac{9271}{504}\nu +
  \frac{65}{18} \nu^2\right) x^2 
  + \left(-\frac{8191}{672}-\frac{583}{24}\nu\right)\pi \,x^{5/2}
  \nonumber \\
  & \quad \quad \quad
  + \left[\frac{6643739519}{69854400}+
  \frac{16}{3}\pi^2-\frac{1712}{105}C -
  \frac{856}{105} \ln (16\,x) \right.
  \nonumber \\
  & \quad \quad \qquad ~
  + \left. \left(-\frac{134543}{7776} +
  \frac{41}{48}\pi^2 
  \right)\nu - \frac{94403}{3024}\nu^2 -
  \frac{775}{324}\nu^3 \right] x^3
  \nonumber \\
  & \quad \quad \quad
  + \left(-\frac{16285}{504} +
  \frac{214745}{1728}\nu + \frac{193385}{3024}\nu^2\right)\pi \,x^{7/2} +
  {\cal O}\left(\frac{1}{c^8}\right) \biggr\}\,.
\end{align}
Here $C=0.577\cdots$ is the Euler constant. This result is fully consistent with black-hole perturbation theory: using it Sasaki \& Tagoshi \cite{Sasa94, TSasa94, TTS96} obtain \eqref{calFx} in the small mass-ratio limit $\nu\rightarrow 0$. The generalization of \eqref{calFx} to arbitrary eccentric (bound) orbits has also been worked out \cite{ABIQ08tail, ABIQ08}.

\section{Waveform of compact binaries}
\label{sec12}

We specify our conventions for the orbital phase and polarization vectors defining the polarization waveforms \eqref{hpc} in the case of a non-spinning compact binary moving on a quasi-circular orbit. If the orbital plane is chosen to be the x-y plane as in Section \ref{sec11}, with the orbital phase $\phi$ measuring the direction of the unit separation vector $\mathbf{n} = \mathbf{x}/r$, then
\begin{equation}\label{n}
\mathbf{n} = \hat{\mathbf{x}}\,\cos{\phi} + \hat{\mathbf{y}}\,\sin{\phi}\,,
\end{equation}
where $\hat{\mathbf{x}}$ and $\hat{\mathbf{y}}$ are the unit directions along x and y. Following \cite{BIWW96, ABIQ04} we choose the polarization vector $\mathbf{P}$ to lie along the x-axis and the observer to be in the y-z plane in the direction
\begin{equation}\label{N}
\mathbf{N} = \hat{\mathbf{y}}\,\sin{i} + \hat{\mathbf{z}}\,\cos{i}\,,
\end{equation}
where $i$ is the orbit's inclination angle. With this choice $\mathbf{P}$ lies along the intersection of the orbital plane with the plane of the sky in the direction of the \textit{ascending node}, \textit{i.e.} that point at which the bodies cross the plane of the sky moving toward the observer. Hence the orbital phase $\phi$ is the angle between the ascending node and the direction of body 1. The rotating orthonormal triad $(\mathbf{n},{\bm \lambda},\hat{\mathbf{z}})$ describing the motion of the binary and used in \eqref{xva} is related to the fixed polarization triad $(\mathbf{N},\mathbf{P},\mathbf{Q})$ by
\begin{subequations}\label{nlz}\begin{align}
\mathbf{n} &= \mathbf{P}\,\cos{\phi} + \bigl( \mathbf{Q}\,\cos{i} +  \mathbf{N}\,\sin{i}\bigr)\,\sin{\phi} \,,\\ {\bm \lambda} &= - \mathbf{P}\,\sin{\phi} +
\bigl( \mathbf{Q}\,\cos{i} +  \mathbf{N}\,\sin{i}\bigr)\,\cos{\phi} \,,\\
\hat{\mathbf{z}} &= - \mathbf{Q}\,\sin{i} + \mathbf{N}\,\cos{i}\,.
\end{align}\end{subequations}

The 3.5PN expression of the orbital phase $\phi$ as function of the orbital frequency or equivalently the $x$-parameter is obtained from the energy balance equation \eqref{balance} in which the binary's conservative center-of-mass energy $E$ and total gravitational-wave flux $\mathcal{F}$ have been obtained in \eqref{Ex}--\eqref{calFx}. For circular orbits the orbital phase is computed from
\begin{equation}\label{phase}
\phi \equiv \int\omega\,\ud t = - \int\frac{\omega}{\mathcal{F}}\frac{\ud E}{\ud\omega}\ud\omega\,.
\end{equation}
Various methods (numerical or analytical) are possible for solving \eqref{phase} given the expressions \eqref{Ex} and \eqref{calFx}. This yields different waveform families all valid at the same 3.5PN order, but which may differ when extrapolated beyond the normal domain of validity of the post-Newtonian expansion, \textit{i.e.} in this case very near the coalescence. Such differences must be taken into account when comparing the post-Newtonian waveforms to numerical results \cite{BCP07}.

It is convenient to perform a change of phase, from the actual orbital phase $\phi$ to the new phase variable
\begin{equation}\label{psi}
\psi=\phi - {2G \,\mathrm{M} \,\omega \over c^3} \ln\left({\omega\over \omega_0}\right)\,,
\end{equation}
where $\mathrm{M}$ is the binary's total mass monopole moment\footnote{The mass monopole $\mathrm{M}$ differs from $m=m_1+m_2$ as it includes the contribution of the gravitational binding energy. At 1PN order it is given for circular orbits by $\mathrm{M} = m [1-\frac{\nu}{2}\gamma]+\mathcal{O}(c^{-4})$.} and $\omega_0 = \frac{1}{4u_0}\mathrm{exp}[\frac{11}{12}-C]$ is related to the constant $u_0\equiv r_0/c$ entering the tail integrals in \eqref{U2}. The logarithmic term in $\psi$ corresponds physically to some spreading of the different frequency components of the wave along the line of sight from the source to the detector, and expresses the tail effect as a small delay in the arrival time of gravitational waves. This effect, although of formal 1.5PN order in \eqref{psi}, represents in fact a very small modulation of the orbital phase: compared to the dominant phase evolution whose order is that of the inverse of 2.5PN radiation reaction, this modulation is of order 4PN and can thus be neglected with the present accuracy.

The spherical harmonic modes of the polarization waveforms can now be obtained at 3PN order using the angular integration formula \eqref{decomp}. We start from the expressions of the wave polarizations $h_+$ and $h_\times$ as functions of the inclination angle $i$ and of the phase $\psi$. We use the known dependence of the spherical harmonics \eqref{harm} on the azimuthal angle. Denoting $h \equiv h_+ - \ui h_\times = h(i,\psi)$ we find that the latter angular integration becomes
\begin{equation}\label{decomp2}
h^{\ell m} = (-\ui)^m\,e^{-\ui \,m \psi}\,\int_0^{2\pi}\!\! \ud\psi' \int_{0}^{\pi} \ud i\,\sin i\,\,h(i,\psi')\,Y^{\,\ell m}_{-2} (i,\psi')\,,
\end{equation}
exhibiting the azimuthal factor $e^{-\ui \,m \psi}$ appropriate for each mode. Let us introduce a normalized mode coefficient $H^{\ell m}$ starting by definition with one at the Newtonian order for the dominant mode having $(\ell,m)=(2,2)$. This means posing
\begin{align}\label{Hhat}
h^{\ell m} = \frac{2 G \,m \,\nu \,x}{R \,c^2} \,\sqrt{\frac{16\pi}{5}}\,H^{\ell m}\,e^{-\ui m \psi}\,.
\end{align}
All the modes have been given in \cite{BFIS08} up to the 3PN order. The dominant mode $(2,2)$, which is primarily needed for the comparison between post-Newtonian calculations and numerical simulations, reads at 3PN order
\begin{align}\label{h22}
H^{22} &= 1 + \left(-\frac{107}{42}+\frac{55}{42}\nu\right) x +2 \pi
\,x^{3/2} \\ & + \left(-\frac{2173}{1512}-\frac{1069}{216}\nu+\frac{2047}{1512}\nu^2\right) x^2 + \left(-\frac{107}{21}\pi+\frac{34}{21}\pi \nu -24 \ui \nu\right) x^{5/2} \nonumber \\ &+ \bigg(\frac{27027409}{646800}-\frac{856}{105}C+\frac{2}{3}\pi ^2 -\frac{428}{105} \ln(16 x)+\frac{428}{105}\ui\,\pi\nonumber \\ &\quad
+\left[-\frac{278185}{33264}+\frac{41}{96}\pi^2\right] \nu -\frac{20261}{2772}\nu^2+\frac{114635}{99792}\nu^3\bigg) x^3 +
\mathcal{O}\left(\frac{1}{c^7}\right)\,.\nonumber
\end{align}

\section{Spin-orbit contributions in the energy and flux} 
\label{sec13}

To successfully detect the gravitational waves emitted by spinning, precessing binaries and to estimate the binary parameters, spin effects should be included in the templates. For maximally spinning compact bodies the spin-orbit coupling (linear in the spins) appears dominantly at the 1.5PN order, while the spin-spin one (which is quadratic) appears at 2PN order. The spin effect on the free motion of a test particle was obtained by Papapetrou \cite{Papa51spin} in the form of a coupling to curvature. Seminal later works by Barker \& O'Connell \cite{BOC75, BOC79} obtained the leading order spin-orbit and spin-spin contributions in the post-Newtonian equations of motion. Based on these works, the spin-orbit and spin-spin terms were obtained in the radiation field \cite{KWW93, K95, Ger99, MVGer05}, enabling the derivation of the orbital phase evolution (the crucial quantity that determines the templates). Finding the 1PN corrections to the leading spin-orbit coupling in both the (translational) equations of motion and radiation field was begun in \cite{OTO98, TOO01} and completed in \cite{FBB06, BBF06}. The result \cite{FBB06} for the equations of motion was confirmed by an alternative derivation based on the ADM-Hamiltonian formalism \cite{DJSspin}.

In Section \ref{sec8} we discussed a covariant formalism for spinning particles \cite{Tulc1, Tulc2, Traut58, Dixon, BI80}. We want now to find a convenient three-dimensional variable for the spin. Restricting ourselves to spin-orbit effects, \textit{i.e.} neglecting $\mathcal{O}(S^2)$, we can write the components of the spin tensor $S\ab_A$ as
\begin{subequations}\label{SmunuA}
\begin{align}
S_A^{0i} &=
-\frac{1}{c\sqrt{-(g)_A}}\,\varepsilon^{ijk}\,u^A_j\,S^A_k
\, , \\ S_A^{ij} &=
-\frac{1}{c\sqrt{-(g)_A}}\,\varepsilon^{ijk}\left[u^A_0\,S^A_k
  +u^A_k \frac{v_A^l}{c}S^A_l\right] \,.
\end{align}
\end{subequations}
We have used the momentum-velocity relation \eqref{pulin} and have taken into account the spin condition \eqref{Su}. A first possibility is to adopt as the basic spin variable the \textit{contravariant} components of the spin covector $S^A_i$ in \eqref{SmunuA}, which are obtained by raising the index by means of the contravariant spatial metric, \textit{viz}
\begin{equation}\label{spinvar}
\mathbf{S}_A=(S_A^i)\quad\text{with}\quad S_A^i\equiv(\gamma^{ij})_AS^A_j\,,
\end{equation}
where $\gamma^{ij}$ is the inverse of the covariant spatial metric $\gamma_{ij}\equiv g_{ij}$ \textit{i.e.} satisfies $\gamma^{ik}\gamma_{kj}=\delta^i_j$. The choice of spin variable \eqref{spinvar} has been adopted in \cite{OTO98, TOO01}. 

However to express final results (to be used in gravitational-wave templates) it is better to use a different set of spin variables characterized by having some \textit{conserved} Euclidean lengths. Such spins with constant Euclidean magnitude will be denoted by $\mathbf{S}^\text{c}_A$. They can be computed in a straightforward way at a given post-Newtonian order in terms of the previous variables \eqref{spinvar}. For instance we find up to 1PN order,
\begin{equation}\label{SAc}
\mathbf{S}^\text{c}_A = \biggl[1+\frac{(U)_A}{c^2}\biggr]\,\mathbf{S}_A-\frac{1}{2c^2}(\mathbf{v}_A\cdot\mathbf{S}_A)\,\mathbf{v}_A + \mathcal{O}\left(\frac{1}{c^4}\right)\,.
\end{equation}
The (regularized) gravitational potential $(U)_A$ is defined by \eqref{UA}. The constant-magnitude spin variable $\mathbf{S}^\text{c}_A$ obeys a spin precession equation which is necessarily of the form
\begin{equation}\label{preceq}
\frac{\ud \mathbf{S}_A^\text{c}}{\ud t} = \bm{\Omega}_A\times
\mathbf{S}_A^\text{c}\,.
\end{equation}
Indeed this equation implies that $\vert\mathbf{S}_A^\text{c}\vert = \mathrm{const}$. The precession angular frequency vector $\bm{\Omega}_A$ for two-body systems has been computed for the leading spin-orbit and spin-spin contributions \cite{K95} and for the 1PN correction to the spin-orbit \cite{FBB06, BBF06, DJSspin}. For two bodies we conveniently use the following combinations (introduced in \cite{K95}) of the two spins:
\begin{subequations}\label{SSigma}
\begin{align}
\mathbf{S}^\text{c} &\equiv \mathbf{S}^\text{c}_1 + \mathbf{S}^\text{c}_2\,,\\ \bm{\Sigma}^\text{c}
&\equiv m\left[\frac{\mathbf{S}^\text{c}_2}{m_2} -
\frac{\mathbf{S}^\text{c}_1}{m_1}\right]\,.
\end{align}
\end{subequations}
Furthermore, recalling the orthonormal triad $\{\mathbf{n},\bm{\lambda},\hat{\mathbf{z}}\}$ used in Section \ref{sec11}, where $\hat{\mathbf{z}}$ is the unit vector in the direction perpendicular to the orbital plane, we denote by $S^\text{c}_\mathrm{z}\equiv\mathbf{S}^\text{c}\cdot \hat{\mathbf{z}}$ and $\Sigma^\text{c}_\mathrm{z}\equiv \bm{\Sigma}^\text{c}\cdot \hat{\mathbf{z}}$ the projections along that perpendicular direction.

The spin-orbit terms have been computed at 1PN order both in the equations of motion and in the radiation field. In the equations of motion they correct the orbital frequency and invariant conserved energy with terms at orders 1.5PN and 2.5PN. In the presence of spins the energy gets modified to $E=E_\text{mono}+E_\text{spin}$ where the monopole part has been obtained in \eqref{Ex} and where the spin terms read
\begin{align}\label{Exspin}
E_\text{spin}=&-\frac{c^2}{2 G\,m}\,\nu\,x\,\left\{ x^{3/2} \left[ \frac{14}{3}S^\text{c}_\mathrm{z} +2\Delta\, \Sigma^\text{c}_\mathrm{z}\right] \right. \nonumber\\
&\left.~+ x^{5/2} \left[
\left(11-\frac{61}{9}\nu\right)S^\text{c}_\mathrm{z}
+\left(3-\frac{10}{3}\nu\right)\Delta\,\Sigma^\text{c}_\mathrm{z}\right]+
\mathcal{O}\left(\frac{1}{c^6}\right)\right\}\,.
\end{align}
We recall that $\Delta\equiv(m_1-m_2)/m$; see the footnote \ref{footnotemass}. This expression is valid for quasi-circular orbits, and we neglect the spin-spin terms. Similarly the gravitational-wave flux will be modified at the same 1.5PN and 2.5PN orders. Posing $\mathcal{F}=\mathcal{F}_\text{mono}+\mathcal{F}_\text{spin}$
where $\mathcal{F}_\text{mono}$ is given by \eqref{calFx}, we find
\begin{align}\label{Fxspin}
\mathcal{F}_\text{spin} =&\frac{32 c^5}{5 G^2\,m^2}\,\nu^2\,x^5\,\left\{ x^{3/2}\left[-4S^\text{c}_\mathrm{z} -\frac{5}{4}\Delta\,\Sigma^\text{c}_\mathrm{z}\right] \right.
\nonumber\\&\left.+x^{5/2}\left[
\left(-\frac{9}{2}+\frac{272}{9}\nu\right)S^\text{c}_\mathrm{z}
+\left(-\frac{13}{16}+\frac{43}{4}\nu\right)\Delta\,\Sigma^\text{c}_\mathrm{z}\right]+
\mathcal{O}\left(\frac{1}{c^6}\right)\right\}\,.
\end{align}

Having in hand the spin contributions in $E$ and $\mathcal{F}$, we can deduce the evolution of the orbital phase from the energy balance equation \eqref{balance}. In absence of precession of the orbital plane, \textit{e.g.} for spins aligned or anti-aligned with the orbital angular momentum, the gravitational-wave phase will reduce to the ``carrier'' phase $\phi_{\rm GW} \equiv 2 \phi$ (keeping only the dominant harmonics), where $\phi$ is the orbital phase which is obtained by integrating the orbital frequency. However, in the general case of non-aligned spins, we must take into account the effect of precession of the orbital plane induced by spin modulations. Then the gravitational-wave phase is given by $\Phi_\mathrm{GW} = \phi_\mathrm{GW} + \delta\phi_\mathrm{GW}$, where the precessional correction $\delta\phi_\mathrm{GW}$ arises from the changing orientation of the orbital plane, and can be computed by standard methods using numerical integration \cite{ACST94}. Thus, the carrier phase $\phi_\mathrm{GW}$ constitutes the main theoretical output to be provided for the gravitational-wave templates, and can directly be obtained numerically from using the integration formula \eqref{phase}. 

%

\bibliography{/home/luc/Articles/ListeRef/ListeRef.bib}

\begin{thebibliography}{100}

\bibitem{E16}
A.~Einstein.
\newblock Sitzber. Preuss. Akad. Wiss. Berlin, 1916.

\bibitem{dS16a}
W.~De~Sitter.
\newblock {\em Mon. Not. Roy. Astron. Soc.}, 76:699, 1916.

\bibitem{dS16b}
W.~De~Sitter.
\newblock {\em Mon. Not. Roy. Astron. Soc.}, 77:155, 1916.

\bibitem{LD17}
H.A. Lorentz and J.~Droste.
\newblock Nijhoff, The Hague, 1937.
\newblock Versl. K. Akad. Wet. Amsterdam {\bf 26}, 392 and 649 (1917).

\bibitem{EIH}
A.~Einstein, L.~Infeld, and B.~Hoffmann.
\newblock The gravitational equations and the problem of motion.
\newblock {\em Ann. Math.}, 39:65--100, 1938.

\bibitem{Fock39}
V.~Fock.
\newblock On motion of finite masses in general relativity.
\newblock {\em J. Phys. (Moscow)}, 1(2):81--116, 1939.

\bibitem{Fock}
V.A. Fock.
\newblock {\em Theory of space, time and gravitation}.
\newblock Pergamon, London, 1959.

\bibitem{PB59}
J.~Plebanski and S.L. Bazanski.
\newblock The general fokker action principle and its application in general
  relativity theory.
\newblock {\em Acta Phys. Polonica}, 18:307--345, 1959.

\bibitem{C65}
S.~Chandrasekhar.
\newblock The post-newtonian equations of hydrodynamics in general relativity.
\newblock {\em Astrophys. J.}, 142:1488--1540, 1965.

\bibitem{CN69}
S.~Chandrasekhar and Y.~Nutku.
\newblock The second post-newtonian equations of hydrodynamics in general
  relativity.
\newblock {\em Astrophys. J.}, 158:55--79, 1969.

\bibitem{CE70}
S.~Chandrasekhar and F.P. Esposito.
\newblock The 5/2-post-newtonian equations of hydrodynamics and radiation
  reaction in general relativity.
\newblock {\em Astrophys. J.}, 160:153--179, 1970.

\bibitem{Ehl80}
J.~Ehlers.
\newblock {\em Ann. N.Y. Acad. Sci.}, 336:279, 1980.

\bibitem{K80a}
G.D. Kerlick.
\newblock {\em Gen. Relativ. Gravit.}, 12:467, 1980.

\bibitem{K80b}
G.D. Kerlick.
\newblock {\em Gen. Relativ. Gravit.}, 12:521, 1980.

\bibitem{Papa51}
A.~Papapetrou.
\newblock Equations of motion in general relativity.
\newblock {\em Proc. Phys. Soc. A}, 64:57--75, 1951.

\bibitem{PapaL81}
A.~Papapetrou and B.~Linet.
\newblock Equation of motion including the reaction of gravitational radiation.
\newblock {\em Gen. Relativ. Gravit.}, 13:335, 1981.

\bibitem{DD81a}
T.~Damour and N.~Deruelle.
\newblock Radiation reaction and angular momentum loss in small angle
  gravitational scattering.
\newblock {\em Phys. Lett. A}, 87:81, 1981.

\bibitem{DD81b}
T.~Damour and N.~Deruelle.
\newblock Lagrangien g{\'e}n{\'e}ralis{\'e} du syst{\`e}me de deux masses
  ponctuelles, {\`a} l'approximation post-post-newtonienne de la relativit{\'e}
  g{\'e}n{\'e}rale.
\newblock {\em C. R. Acad. Sc. Paris}, 293:537, 1981.

\bibitem{D83}
T.~Damour.
\newblock Gravitational radiation reaction in the binary pulsar and the
  quadrupole formula controvercy.
\newblock {\em Phys. Rev. Lett.}, 51:1019--1021, 1983.

\bibitem{D83houches}
T.~Damour.
\newblock Gravitational radiation and the motion of compact bodies.
\newblock In N.~Deruelle and T.~Piran, editors, {\em Gravitational Radiation},
  pages 59--144, Amsterdam, 1983. North-Holland Company.

\bibitem{TFMc79}
J.H. Taylor, L.A. Fowler, and P.M. McCulloch.
\newblock Measurements of general relativistic effects in the binary pulsar
  psr1913+16.
\newblock {\em Nature}, 277:437--440, 1979.

\bibitem{TW82}
J.H. Taylor and J.M. Weisberg.
\newblock A new test of general relativity: Gravitational radiation and the
  binary pulsar psr 1913+16.
\newblock {\em Astrophys. J.}, 253:908--920, 1982.

\bibitem{T93}
J.H. Taylor.
\newblock Pulsar timing and relativistic gravity.
\newblock {\em Class. Quant. Grav.}, 10:167--174, 1993.

\bibitem{3mn}
C.~Cutler, T.A. Apostolatos, L.~Bildsten, L.S. Finn, E.E. Flanagan,
  D.~Kennefick, D.M. Markovic, A.~Ori, E.~Poisson, G.J. Sussman, and K.S.
  Thorne.
\newblock The last three minutes: Issues in gravitational-wave measurements of
  coalescing compact binaries.
\newblock {\em Phys. Rev. Lett.}, 70:2984--2987, 1993.

\bibitem{CFPS93}
C.~Cutler, L.S. Finn, E.~Poisson, and G.J. Sussman.
\newblock Gravitational radiation from a particle in circular orbit around a
  black hole. ii. numerical results for the nonrotating case.
\newblock {\em Phys. Rev. D}, 47:1511--1518, 1993.

\bibitem{FCh93}
L.S. Finn and D.F. Chernoff.
\newblock Observing binary inspiral in gravitational radiation: One
  interferometer.
\newblock {\em Phys. Rev. D}, 47:2198--2219, 1993.

\bibitem{CF94}
C.~Cutler and E.E. Flanagan.
\newblock Gravitational waves from merging compact binaries: How accurately can
  one extract the binary's parameters from the inspiral waveform?
\newblock {\em Phys. Rev. D}, 49:2658--2697, 1994.

\bibitem{TNaka94}
H.~Tagoshi and T.~Nakamura.
\newblock Gravitational waves from a point particle in circular orbit around a
  black hole: Logarithmic terms in the post-newtonian expansion.
\newblock {\em Phys. Rev. D}, 49:4016--4022, 1994.

\bibitem{P95}
E.~Poisson.
\newblock Gravitational radiation from a particle in circular orbit around a
  black-hole. vi. accuracy of the post-newtonian expansion.
\newblock {\em Phys. Rev. D}, 52:5719--5723, 1995.
\newblock Erratum Phys. Rev. D {\bf 55}, 7980, (1997).

\bibitem{Pe64}
P.C. Peters.
\newblock Gravitational radiation and the motion of two point masses.
\newblock {\em Phys. Rev.}, 136:B1224--B1232, 1964.

\bibitem{AD75}
J.L. Anderson and T.C. DeCanio.
\newblock {\em Gen. Relativ. Gravit.}, 6:197, 1975.

\bibitem{Rend92}
A.D. Rendall.
\newblock On the definition of post-newtonian approximations.
\newblock {\em Proc. R. Soc. Lond. A}, 438:341, 1992.

\bibitem{PB02}
O.~Poujade and L.~Blanchet.
\newblock Post-newtonian approximation for isolated systems calculated by
  matched asymptotic expansions.
\newblock {\em Phys. Rev. D}, 65:124020, 2002.

\bibitem{FS83}
T.~Futamase and B.F. Schutz.
\newblock {\em Phys. Rev. D}, 28:2363, 1983.

\bibitem{F83}
T.~Futamase.
\newblock {\em Phys. Rev. D}, 28:2373, 1983.

\bibitem{BD86}
L.~Blanchet and T.~Damour.
\newblock Radiative gravitational fields in general relativity i. general
  structure of the field outside the source.
\newblock {\em Phil. Trans. Roy. Soc. Lond. A}, 320:379--430, 1986.

\bibitem{BD88}
L.~Blanchet and T.~Damour.
\newblock Tail transported temporal correlations in the dynamics of a
  gravitating system.
\newblock {\em Phys. Rev. D}, 37:1410, 1988.

\bibitem{B93}
L.~Blanchet.
\newblock Time asymmetric structure of gravitational radiation.
\newblock {\em Phys. Rev. D}, 47:4392--4420, 1993.

\bibitem{BD92}
L.~Blanchet and T.~Damour.
\newblock Hereditary effects in gravitational radiation.
\newblock {\em Phys. Rev. D}, 46:4304--4319, 1992.

\bibitem{BuTh70}
W.L. Burke and K.S. Thorne.
\newblock Gravitational radiation damping.
\newblock In M.~Carmeli, S.I. Fickler, and L.~Witten, editors, {\em
  Relativity}, pages 209--228, New York and London, 1970. Plenum Press.

\bibitem{Bu71}
W.L. Burke.
\newblock Gravitational radiation damping of slowly moving systems calculated
  using matched asymptotic expansions.
\newblock {\em J. Math. Phys.}, 12:401, 1971.

\bibitem{BD89}
L.~Blanchet and T.~Damour.
\newblock Post-newtonian generation of gravitational waves.
\newblock {\em Annales Inst. H. Poincar{\'e} Phys. Th\'eor.}, 50:377--408,
  1989.

\bibitem{DI91a}
T.~Damour and B.~R. Iyer.
\newblock Postnewtonian generation of gravitational waves. 2. the spin moments.
\newblock {\em Annales Inst. H. Poincar{\'e}, Phys. Th\'eor.}, 54:115--164,
  1991.

\bibitem{B98mult}
L.~Blanchet.
\newblock On the multipole expansion of the gravitational field.
\newblock {\em Class. Quant. Grav.}, 15:1971--1999, 1998.

\bibitem{BFN05}
L.~Blanchet, G.~Faye, and Samaya Nissanke.
\newblock Structure of the post-newtonian expansion in general relativity.
\newblock {\em Phys. Rev. D}, 72:044024, 2005.

\bibitem{WW96}
C.M. Will and A.G. Wiseman.
\newblock Gravitational radiation from compact binary systems: Gravitational
  waveforms and energy loss to second post-newtonian order.
\newblock {\em Phys. Rev. D}, 54:4813--4848, 1996.

\bibitem{PW00}
M.E. Pati and C.M. Will.
\newblock Post-newtonian gravitational radiation and equations of motion via
  direct integration of the relaxed einstein equations: Foundations.
\newblock {\em Phys. Rev. D}, 62:124015, 2000.

\bibitem{PW02}
M.E. Pati and C.M. Will.
\newblock Post-newtonian gravitational radiation and equations of motion via
  direct integration of the relaxed einstein equations. ii. two-body equations
  of motion to second post-newtonian order, and radiation-reaction to 3.5
  post-newtonian order.
\newblock {\em Phys. Rev. D}, 65:104008, 2002.

\bibitem{Th80}
K.S. Thorne.
\newblock Multipole expansions of gravitational radiation.
\newblock {\em Rev. Mod. Phys.}, 52:299--339, 1980.

\bibitem{BBM62}
H.~Bondi, M.G.J. van~der Burg, and A.W.K. Metzner.
\newblock Gravitational waves in general relativity vii. waves from
  axi-symmetric isolated systems.
\newblock {\em Proc. R. Soc. London, Ser. A}, 269:21, 1962.

\bibitem{B87}
L.~Blanchet.
\newblock Radiative gravitational fields in general relativity. 2. asymptotic
  behaviour at future null infinity.
\newblock {\em Proc. Roy. Soc. Lond. A}, 409:383--399, 1987.

\bibitem{BCP07}
A.~Buonanno, G.~B. Cook, and F.~Pretorius.
\newblock {\em Phys. Rev. D}, 75:124018, 2007.

\bibitem{K07}
L.E. Kidder.
\newblock Using full information when computing modes of post-newtonian
  waveforms from inspiralling compact binaries in circular orbits.
\newblock {\em Phys. Rev. D}, 77:044016, 2008.

\bibitem{BFIS08}
L.~Blanchet, G.~Faye, B.~R. Iyer, and Siddhartha Sinha.
\newblock The third post-newtonian gravitational wave polarisations and
  associated spherical harmonic modes for inspiralling compact binaries in
  quasi-circular orbits.
\newblock {\em Class. Quant. Grav.}, 25:165003, 2008.

\bibitem{B98tail}
L.~Blanchet.
\newblock Gravitational-wave tails of tails.
\newblock {\em Class. Quant. Grav.}, 15:113--141, 1998.
\newblock Erratum \textit{Class. Quant. Grav.}, 22:3381, 2005.

\bibitem{Chr91}
D.~Christodoulou.
\newblock Nonlinear nature of gravitation and gravitational-wave experiments.
\newblock {\em Phys. Rev. Lett.}, 67:1486--1489, 1991.

\bibitem{Th92}
K.S. Thorne.
\newblock Gravitational-wave bursts with memory: The christodoulou effect.
\newblock {\em Phys. Rev. D}, 45:520, 1992.

\bibitem{WW91}
A.G. Wiseman and C.M. Will.
\newblock Christodoulou's nonlinear gravitational-wave memory: Evaluation in
  the quadrupole approximation.
\newblock {\em Phys. Rev. D}, 44:R2945--R2949, 1991.

\bibitem{Tulc1}
W.~Tulczyjew.
\newblock {\em Bull. Acad. Polon. Sci.}, 5:279, 1957.

\bibitem{Tulc2}
W.~Tulczyjew.
\newblock {\em Acta Phys. Polon.}, 18:37, 1959.

\bibitem{Traut58}
A.~Trautman.
\newblock Lectures on general relativity.
\newblock {\em Gen. Relat. Grav.}, 34:721, 2002.
\newblock reprinted from lectures delivered in 1958.

\bibitem{Dixon}
W.G. Dixon.
\newblock In J.~Ehlers, editor, {\em Isolated systems in general relativity},
  page 156, Amsterdam, 1979. North Holland.

\bibitem{BI80}
I.~Bailey and W.~Israel.
\newblock {\em Ann. Phys.}, 130:188, 1980.

\bibitem{Papa51spin}
A.~Papapetrou.
\newblock Spinning test-particles in general relativity. i.
\newblock {\em Proc. R. Soc. London A}, 209:248, 1951.

\bibitem{K95}
L.E. Kidder.
\newblock Coalescing binary systems of compact objects to 5/2-post-newtonian
  order. v. spin effects.
\newblock {\em Phys. Rev. D}, 52:821--847, 1995.

\bibitem{Hadamard}
J.~Hadamard.
\newblock {\em Le probl\`eme de Cauchy et les \'equations aux d\'eriv\'ees
  partielles lin\'eaires hyperboliques}.
\newblock Hermann, Paris, 1932.

\bibitem{Schwartz}
L.~Schwartz.
\newblock {\em Th\'eorie des distributions}.
\newblock Hermann, Paris, 1978.

\bibitem{BFreg}
L.~Blanchet and G.~Faye.
\newblock Hadamard regularization.
\newblock {\em J. Math. Phys.}, 41:7675--7714, 2000.

\bibitem{BFeom}
L.~Blanchet and G.~Faye.
\newblock General relativistic dynamics of compact binaries at the third
  post-newtonian order.
\newblock {\em Phys. Rev. D}, 63:062005, 2001.

\bibitem{tHooft}
G.~'t~Hooft and M.~Veltman.
\newblock {\em Nucl. Phys.}, B44:139, 1972.

\bibitem{Bollini}
C.~G. Bollini and J.~J. Giambiagi.
\newblock {\em Phys. Lett. B}, 40:566, 1972.

\bibitem{DJSdim}
T.~Damour, P.~Jaranowski, and G.~Sch\"afer.
\newblock Dimensional regularization of the gravitational interaction of point
  masses.
\newblock {\em Phys. Lett. B}, 513:147--155, 2001.

\bibitem{BDE04}
L.~Blanchet, T.~Damour, and G.~Esposito-Far{\`e}se.
\newblock Dimensional regularization of the third post-newtonian dynamics of
  point particles in harmonic coordinates.
\newblock {\em Phys. Rev. D}, 69:124007, 2004.

\bibitem{BDEI04}
L.~Blanchet, T.~Damour, G.~Esposito-Far{\`e}se, and B.~R. Iyer.
\newblock Gravitational radiation from inspiralling compact binaries completed
  at the third post-newtonian order.
\newblock {\em Phys. Rev. Lett.}, 93:091101, 2004.

\bibitem{ABF01}
V.C. de~Andrade, L.~Blanchet, and G.~Faye.
\newblock Third post-newtonian dynamics of compact binaries: Noetherian
  conserved quantities and equivalence between the harmonic-coordinate and
  adm-hamiltonian formalisms.
\newblock {\em Class. Quant. Grav.}, 18:753--778, 2001.

\bibitem{IFA01}
Y.~Itoh, T.~Futamase, and H.~Asada.
\newblock Equation of motion for relativistic compact binaries with the strong
  field point particle limit: The second and half post-newtonian order.
\newblock {\em Phys. Rev. D}, 63:064038, 2001.

\bibitem{itoh1}
Y.~Itoh and T.~Futamase.
\newblock {\em Phys. Rev. D}, 68:121501(R), 2003.

\bibitem{itoh2}
Y.~Itoh.
\newblock {\em Phys. Rev. D}, 69:064018, 2004.

\bibitem{JaraS98}
P.~Jaranowski and G.~Sch\"afer.
\newblock Third post-newtonian higher order adm hamilton dynamics for two-body
  point-mass systems.
\newblock {\em Phys. Rev. D}, 57:7274--7291, 1998.

\bibitem{JaraS99}
P.~Jaranowski and G.~Sch\"afer.
\newblock Binary black-hole problem at the third post-newtonian approximation
  in the orbital motion: Static part.
\newblock {\em Phys. Rev. D}, 60:124003--1--12403--7, 1999.

\bibitem{BIJ02}
L.~Blanchet, B.~R. Iyer, and B.~Joguet.
\newblock Gravitational waves from inspiralling compact binaries: Energy flux
  to third post-newtonian order.
\newblock {\em Phys. Rev. D}, 65:064005, 2002.
\newblock Erratum \textit{Phys. Rev. D}, 71:129903(E), 2005.

\bibitem{BFIJ02}
L.~Blanchet, G.~Faye, B.~R. Iyer, and B.~Joguet.
\newblock Gravitational-wave inspiral of compact binary systems to 7/2
  post-newtonian order.
\newblock {\em Phys. Rev. D}, 65:061501(R), 2002.
\newblock Erratum \textit{Phys. Rev. D}, 71:129902(E), 2005.

\bibitem{Sasa94}
M.~Sasaki.
\newblock Post-newtonian expansion of the ingoing-wave regge-wheeler function.
\newblock {\em Prog. Theor. Phys.}, 92:17--36, 1994.

\bibitem{TSasa94}
H.~Tagoshi and M.~Sasaki.
\newblock Post-newtonian expansion of gravitational-waves from a particle in
  circular orbit around a schwarzschild black-hole.
\newblock {\em Prog. Theor. Phys.}, 92:745--771, 1994.

\bibitem{TTS96}
T.~Tanaka, H.~Tagoshi, and M.~Sasaki.
\newblock Gravitational waves by a particle in circular orbit around a
  schwarzschild black hole.
\newblock {\em Prog. Theor. Phys.}, 96:1087--1101, 1996.

\bibitem{ABIQ08tail}
K.G. Arun, L.~Blanchet, B.~R. Iyer, and M.~S. Qusailah.
\newblock Tail effects in the 3pn gravitational wave energy flux of compact
  binaries in quasi-elliptical orbits.
\newblock {\em Phys. Rev. D}, 77:064034, 2008.

\bibitem{ABIQ08}
K.G. Arun, L.~Blanchet, B.~R. Iyer, and M.~S. Qusailah.
\newblock Inspiralling compact binaries in quasi-elliptical orbits: The
  complete 3pn energy flux.
\newblock {\em Phys. Rev. D}, 77:064035, 2008.

\bibitem{BIWW96}
L.~Blanchet, B.~R. Iyer, C.~M. Will, and A.~G. Wiseman.
\newblock Gravitational wave forms from inspiralling compact binaries to
  second-post-newtonian order.
\newblock {\em Class. Quant. Grav.}, 13:575--584, 1996.

\bibitem{ABIQ04}
K.G. Arun, L.~Blanchet, B.~R. Iyer, and M.~S. Qusailah.
\newblock The 2.5pn gravitational wave polarisations from inspiralling compact
  binaries in circular orbits.
\newblock {\em Class. Quant. Grav.}, 21:3771, 2004.
\newblock Erratum \textit{Class. Quant. Grav.}, 22:3115, 2005.

\bibitem{BOC75}
B.M. Barker and R.F. O'Connell.
\newblock {\em Phys. Rev. D}, 12:329, 1975.

\bibitem{BOC79}
B.M. Barker and R.F. O'Connell.
\newblock {\em Gen. Relativ. Gravit.}, 11:149, 1979.

\bibitem{KWW93}
L.E. Kidder, C.M. Will, and A.G. Wiseman.
\newblock Spin effects in the inspiral of coalescing compact binaries.
\newblock {\em Phys. Rev. D}, 47:R4183--R4187, 1993.

\bibitem{Ger99}
L.~Gergely.
\newblock Spin-spin effects in radiating compact binaries.
\newblock {\em Phys. Rev. D}, 61:024035, 1999.

\bibitem{MVGer05}
B.~Mik\'oczi, M.~Vas\'uth, and L.~Gergely.
\newblock Self-interaction spin effects in inspiralling compact binaries.
\newblock {\em Phys. Rev. D}, 71:124043, 2005.

\bibitem{OTO98}
B.J. Owen, H.~Tagoshi, and A.~Ohashi.
\newblock Nonprecessional spin-orbit effects on gravitational waves from
  inspiraling compact binaries to second post-newtonian order.
\newblock {\em Phys. Rev. D}, 57:6168--6175, 1998.

\bibitem{TOO01}
H.~Tagoshi, A.~Ohashi, and B.J. Owen.
\newblock Gravitational field and equations of motion of spinning compact
  binaries to 2.5-post-newtonian order.
\newblock {\em Phys. Rev. D}, 63:044006, 2001.

\bibitem{FBB06}
G.~Faye, L.~Blanchet, and A.~Buonanno.
\newblock Higher-order spin effects in the dynamics of compact binaries i.
  equations of motion.
\newblock {\em Phys. Rev. D}, 74:104033, 2006.

\bibitem{BBF06}
L.~Blanchet, A.~Buonanno, and G.~Faye.
\newblock Higher-order spin effects in the dynamics of compact binaries ii.
  radiation field.
\newblock {\em Phys. Rev. D}, 74:104034, 2006.
\newblock Erratum \textit{Phys. Rev. D}, 75:049903, 2007.

\bibitem{DJSspin}
T.~Damour, Piotr Jaranowski, and Gerhard Sch{\"a}fer.
\newblock Hamiltonian of two spinning compact bodies with next-to-leading order
  gravitational spin-orbit coupling.
\newblock {\em Phys. Rev. D}, 77:064032, 2008.

\bibitem{ACST94}
T.A. Apostolatos, C.~Cutler, G.J. Sussman, and K.S. Thorne.
\newblock {\em Phys. Rev. D}, 49:6274, 1994.

\end{thebibliography}

\end{document}